\documentclass[twocolumn]{aastex61}

\usepackage{natbib}

\newcommand{\be}{\begin{equation}}
\newcommand{\ee}{\end{equation}}

\newcommand{\rxj}{RX\,J0806.4--4123 }

\def\be{\begin{equation}}
\def\ee{\end{equation}}
\def\ba{\begin{eqnarray}}
\def\ea{\end{eqnarray}}
\def\m{\mathrm}

\def\Fscat{F_{\mathrm{scat}}}
\def\Fin{F_{\mathrm{in}}}
\def\Ftot{F_{\mathrm{tot}}}
\def\Lin{L_{\mathrm{in}}}

\def\Lstar{L_\ast}
\def\Lscat{L_{\mathrm{scat}}}

\def\OmegaK{\Omega_{\mathrm{K}}}
\def\rin{r_{\mathrm{in}}}
\def\rout{r_{\mathrm{out}}}
\def\rau{r_\mathrm{AU}}

\def\m{\mathrm}
\def\p{\propto}

\def\cs{c_{\mathrm{s}}}
\def\Caliskan{\c{C}al{\i}\c{s}kan~}
\def\ergpers{erg s$^{-1}$}

\shorttitle{Extended NIR emission around RX\,J0806.4--4123}
\shortauthors{Posselt et al.}

\begin{document}
 
\title{Discovery of extended infrared emission around the neutron star RX\,J0806.4--4123\footnote{Based on observations made with the NASA/ESA Hubble Space Telescope, obtained at the Space Telescope Science Institute, which is operated by the Association of Universities for Research in Astronomy, Inc., under NASA contract NAS 5-26555. These observations are associated with program GO-14745.}\footnote{Based on observations obtained at the Gemini Observatory, which is operated by the Association of Universities for Research in Astronomy, Inc., under a cooperative agreement with the NSF on behalf of the Gemini partnership: the National Science Foundation (United States), the National Research Council (Canada), CONICYT (Chile), Ministerio de Ciencia, Tecnolog\'{i}a e Innovaci\'{o}n Productiva (Argentina), and Minist\'{e}rio da Ci\^{e}ncia, Tecnologia e Inova\c{c}\~{a}o (Brazil).}}

\author{B. Posselt}
\correspondingauthor{B.Posselt}
\affil{Department of Astronomy \& Astrophysics, Pennsylvania State University, 525 Davey Lab,University Park, PA 16802, USA}
\email{posselt@psu.edu}

\author{G. G. Pavlov}
\affil{Department of Astronomy \& Astrophysics, Pennsylvania State University, 525 Davey Lab,University Park, PA 16802, USA}

\author{\"{U}. Ertan}
\affil{Sabanc\i\ University, 34956, Orhanl\i\, Tuzla, \.Istanbul, Turkey}

\author{S. \Caliskan}
\affil{Sabanc\i\ University, 34956, Orhanl\i\, Tuzla, \.Istanbul, Turkey}

\author{K. L. Luhman}
\affil{Department of Astronomy \& Astrophysics, Pennsylvania State University, 525 Davey Lab,University Park, PA 16802, USA}
\affil{Center for Exoplanets and Habitable Worlds,
The Pennsylvania State University, University Park, PA 16802, USA}

\author{C. C. Williams}
\affil{Steward Observatory, University of Arizona, 933 North Cherry Avenue, Tucson, AZ 85721, USA}

\begin{abstract}
Following up on a faint detection of a near-infrared (NIR) source at the position of the X-ray thermal isolated neutron star RX\,J0806.4--4123, we present new \emph{Hubble Space Telescope} (\emph{HST}) observations in the $H$-band.
The NIR source is unambiguously detected with a Vega magnitude of $23.7\pm 0.2$ (flux density of $0.40 \pm 0.06$\,$\mu$Jy at $\lambda=1.54\,\mu$m).
The source position is coincident with the neutron star position, and the implied NIR flux is strongly in excess of what one would expect from an extrapolation of the optical-UV spectrum of RX\,J0806.4--4123. 
The NIR source is extended with a size of at least $0\farcs{8}$ and shows some asymmetry. 
The conservative upper limit on the flux contribution of a point source is 50\%.
Emission from gas and dust in the ambient diffuse interstellar medium can be excluded as cause for the extended emission.
The source parameters are consistent with an interpretation as either the first NIR-only detected pulsar wind nebula or the first resolved disk around an isolated neutron star.   
\end{abstract}


\section{Introduction} \label{intro}

The X-ray pulsar RX\,J0806.4--4123 belongs to a group of seven nearby ($<1$\,kpc), X-ray thermal isolated neutron stars (XTINSs, {also dubbed the ``Magnificent Seven''}) whose defining properties are: the lack of detected non-thermal X-ray emission, radio-quietness, relatively large pulse periods ($3-11$\,s) and inferred dipole magnetic fields on the order of $\sim 10^{13}$\,G.
The observational properties of the XTINSs (e.g., \citealt{Haberl2013,Kaplan2009}) place this neutron star  population between those of the  rotation-powered pulsars and the magnetars. Since their X-ray luminosities exceed their spin-down energies by a factor of at least 10, an explanation of the XTINS properties requires an additional energy source other than the energy loss of a rotating dipole.
Currently, it is commonly assumed that the most promising model for the heating of magnetars and XTINSs is the decay of their large magnetic fields. \citet{Vigano2013} presented simulations of a magnetothermal evolution model capable of  unifying the diverse neutron star populations. In this model, magnetars and XTINSs are hot because they are heated by the decay of their large magnetic fields.
In an alternative model, the observational properties of the XTINSs and magnetars are explained by formation of and influence by supernova fallback disks (e.g., \citealt{Ertan2014}). However, the fallback disk model alone cannot explain the giant outbursts of magnetars. Acceptance of this model is also hampered by the lack of fallback disk detections. So far, there is only one example of a 
{possible}
(passive) fallback disk around the 3.9 kpc distant anomalous X-ray pulsar 4U 0142+61, detected with \emph{Spitzer} IRAC by \citet{Wang2006}. \citet{Ertan2007} reported that an active irradiated disk model can also reproduce the observed optical and infrared fluxes of 4U 0142+61.\\

RX\,J0806.4--4123 has the second-longest period among the XTINSs, $P=11.37$\,s. Its spin-down power is $\dot{E}=1.6 \times 10^{30}$\,erg\,s$^{-1}$, its inferred dipole magnetic field is $B=2.5 \times 10^{13}$\,G, and its total X-ray luminosity at a distance of 250\,pc is $L_X=2.6 \times 10^{31}$\,erg\,s$^{-1}$  \citep{Kaplan2009,Haberl2007,Posselt2007,Haberl2004}.
Previously, we used the Very Large Telescope (VLT) in the near-infrared (NIR) to carry out a direct-imaging search for substellar companions around  isolated XTINSs \citep{Posselt2009}, and we noticed a slight $H$-band flux enhancement at the location of RX\,J0806.4--4123. Motivated by an additional {\sl Herschel} 160\,$\mu$m detection very close to the neutron star position \citep{Posselt2014}, we followed up on the NIR source with the Gemini telescope.
We indeed found indications of the VLT source in another NIR band. The two independent faint flux enhancements together, $H_{\rm VLT} \approx 23.3 \pm 0.5 (1\sigma)$ and $J_{\rm Gemini} \approx 24.8 \pm 0.5 (1\sigma)$, resulted in a combined NIR detection ($H$ and $J$ bands) significance of $3.1\sigma$ \citep{Posselt2016}. 
From the UV-optical spectral slope (measured by \citealt{Kaplan2011}), one would expect the XTINS to have a NIR magnitude on the order of 28, much fainter than the NIR detection. Possible interpretations of the NIR flux enhancements could be a substellar companion, a fallback disk or even a very unusual pulsar wind nebula.
In this paper, we report on the results of recently obtained {\sl{Hubble Space Telescope}} (\emph{HST}) NIR imaging follow-up observations that were carried out to confirm the previous weak NIR detection.  

\section{Data reduction and Results} \label{datared}

\subsection{Hubble Space Telescope data} \label{obshubble}
\begin{figure}[t]
\includegraphics[width=8.5cm]{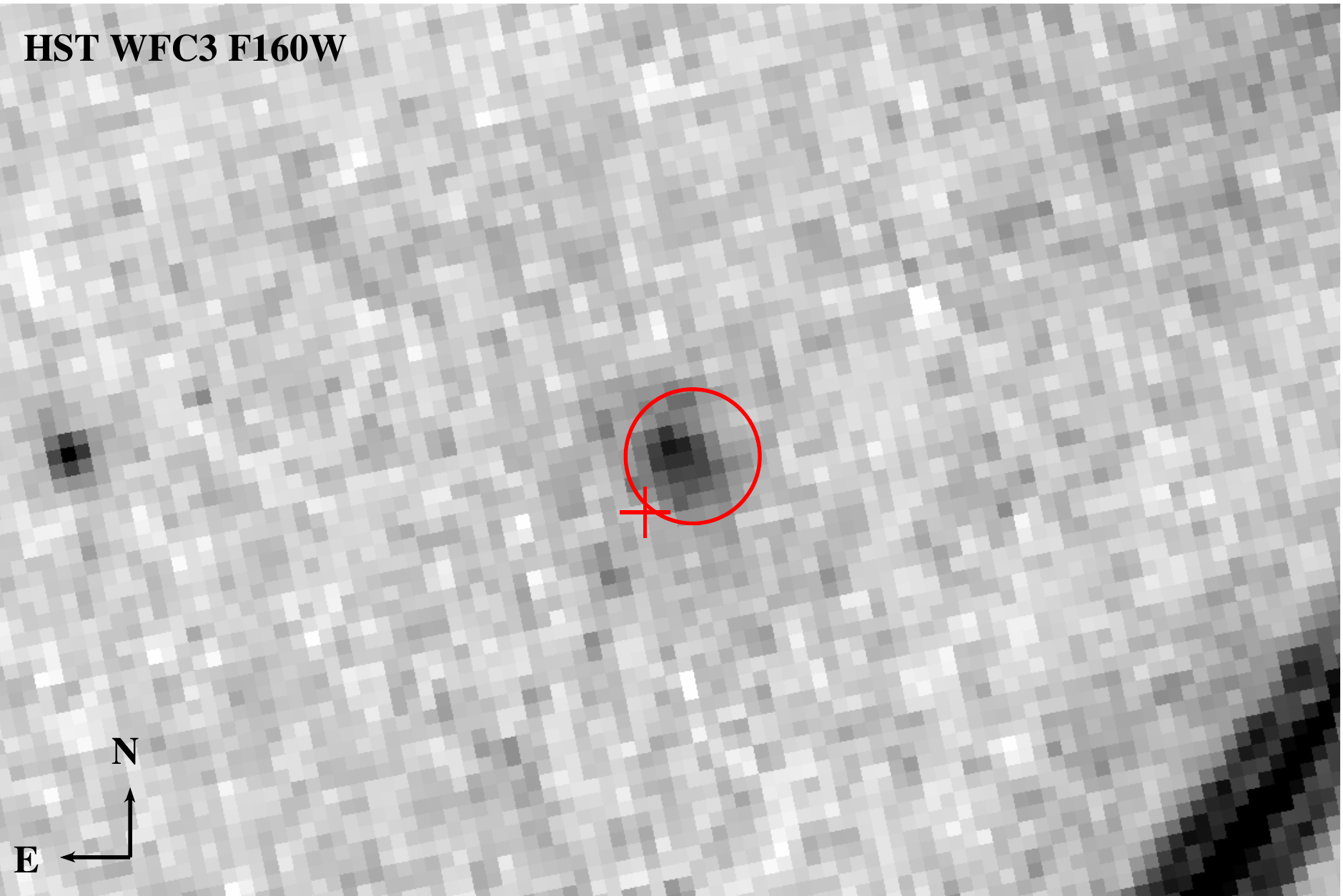}
\caption{The 2016 \emph{HST} F160W image ($6\arcsec \times 4\arcsec$) at the position of RX\,J0806.4--4123. The red circle ($r=0\farcs{3}$) is centered on the previous (2009 and 2010) \emph{HST} F475W detection (STmag$=27.9$) of the neutron star by \citet{Kaplan2011}, and the red cross marks the 2002 {\sl Chandra} position that has a positional uncertainty of  $0\farcs{6}$ \citep{Haberl2004}. 
\label{CXOpos}}
\end{figure}

\begin{figure*}[!]
\gridline{\fig{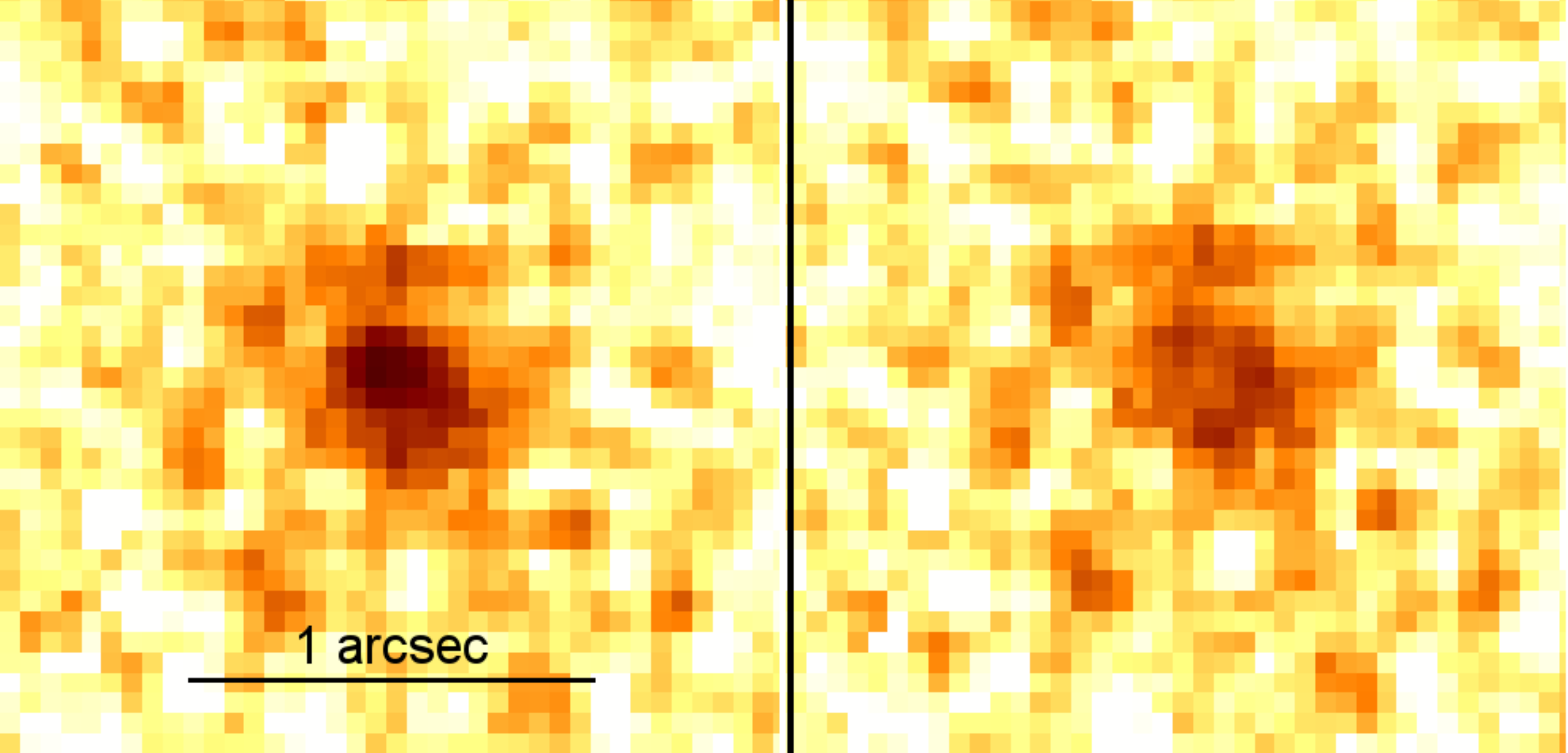}{0.49\textwidth}{The target, STmag=27.4, R$_{\rm maxmin}=0.63$} 
\fig{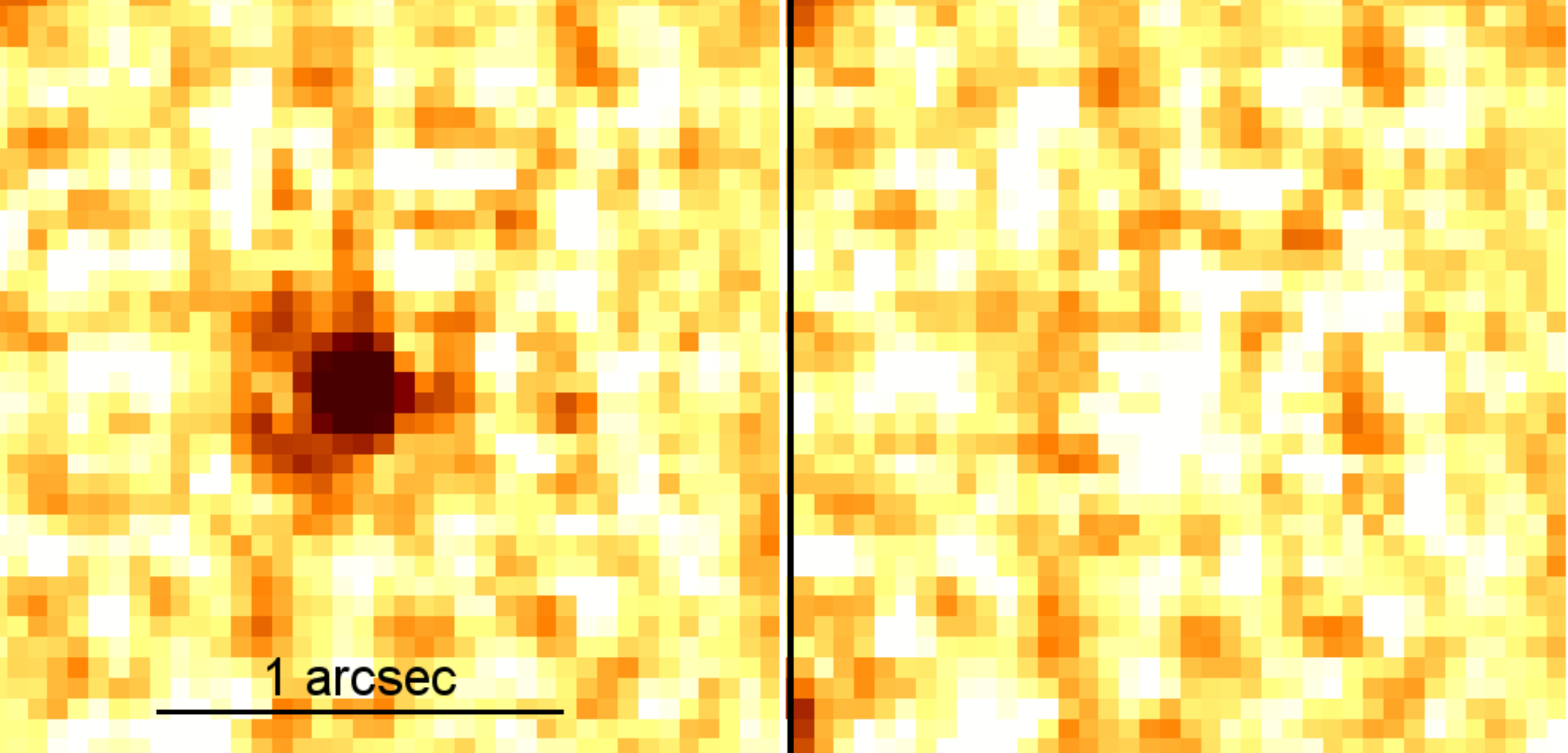}{0.49\textwidth}{s4, STmag$=27.5$, R$_{\rm maxmin}=0.17$}}
\caption{The comparison of images before (only sky subtraction) and after PSF (and sky) subtraction. We show the target (left) and a similarly faint source (right). The images employ the same limits, logarithmic scale and color map. R$_{\rm maxmin}$ is defined as the ratio of the maximum$-$minimum in the PSF-subtracted image (i.e., the residual range; right subpanels) to the maximum$-$minimum in the image without PSF subtraction (left subpanels) and was obtained from image statistics in the inner $0\farcs{7} \times 0\farcs{7}$ regions. 
For the target,  R$_{\rm maxmin}$ is about a factor 4 larger than it is for s4, underlining that there is substantial ``residual flux'' after PSF subtraction in case of the target.    
For other reference source examples in the same observation, see the Appendix, Figures~\ref{fig:psfsubtract1} and \ref{fig:psfsubtract2}.
 \label{fig:targetpsfsub}}
\end{figure*}

We obtained NIR images (program GO-14745) of \rxj with the \emph{HST} Wide Field Camera 3 (WFC3; \citealt{Kimble2008}) on 2016 September 11 (MJD 57642.98), employing the F160W filter with a pivot wavelength of $1.537\mu$m.
The total exposure time was 5417\,s. We used the WIDE-6 POS-TARGs dither pattern (optimizes the subsampling of the pixels; \emph{HST} Instrument Science Report 2016-14\footnote{ISR 2016-14 Supplemental Dither Patterns for WFC3/IR, J. Anderson;  \url{http://www.stsci.edu/hst/wfc3/documents/ISRs}})
for our 6 individual exposures in MULTIACCUM mode (linear SPARS timing sequence).
We stacked and processed our images within PyRAF using AstroDrizzle (version 2.1.8) of the DrizzlePAC software \citep{Gonzaga2012,Fruchter2010}. Since our target source is very faint, we used the inverse variance map (IVM) weighting scheme for the final combination of the data \citep{Gonzaga2012}. 
We experimented with the drizzle parameters for a balance of spatial resolution and sampling noise. We chose a final pixel scale of $0\farcs{0642}$ pixel$^{-1}$ and a pixel fraction of 0.7.   
We registered our astrometry to the \emph{Gaia} data release 1 \citep{Gaia2016b,Gaia2016a}; employing the common 117 reference sources in the field of view, the rms of our astrometry fit is $0\farcs{019}$.\\ 

\label{HSTnir}
RX\,J0806.4--4123 was observed in X-rays in 2002, 2005, and 2015 with {\sl Chandra}. From these observations the proper motion was found to be rather small, $<50$\,mas/yr, (\citealt{Haberl2004,Motch2009}; Posselt et al., {in prep.})
RX\,J0806.4--4123 was also observed and detected with \emph{HST} in the UV (ACS/SBC F140LP; STmag$=23.61 \pm 0.11$) and in the optical (ACS/WFC F475W, STmag$=27.92\pm 0.22$) in 2009 and 2010 \citep{Kaplan2011}. 
The neutron star's position in X-rays and in particular in the optical F475W band are consistent with an unambiguous \emph{HST} detection in the F160W band, see Figure~\ref{CXOpos}.\\

To measure the source flux we used aperture measurements with a sky annulus for the background and noise estimate. We used the \texttt{daophot} and \texttt{photcal} IRAF packages. In order to account for systematic errors and an apparent small extension, we measured values for four different aperture radii ($r=0\farcs{4},0\farcs{5},0\farcs{6}, 0\farcs{7}$), six different sky annuli (inner radius from $r=0\farcs{96}$ to $1\farcs{6}$ with different widths ranging from $\triangle r=0\farcs{6}$ to $1\farcs{5}$ where the latter is limited by a nearby diffraction spike).
We also checked 10 slightly different aperture center positions (maximal difference $\sim 0\farcs{3}$). The largest systematic effect on the measured magnitude came from increasing the aperture radius from $0\farcs{4}$ to $0\farcs{6}$ (up to $\sim 0.3$ brighter magnitude), indicating that the source may have an extension of up to $r=1\farcs{2}$. 
The effect of changing the aperture center position reached $\sim 0.1$ in magnitude. The magnitude  effect of different sky annuli was much smaller ($\sim 0.03$).
Considering these different measurements, we obtained the following median values and $1\sigma$ uncertainties (including systematic errors), a Vega magnitude of $23.7\pm 0.2$, an ST magnitude of $27.1 \pm 0.2$, corresponding to a flux density of $0.40 \pm 0.06$\,$\mu$Jy.\\

To test the hypothesis that the source is extended, we subtracted the point spread function (PSF) from the emission.
In a drizzled image, pixels are correlated due to the drizzling process, and different PSFs are expected for individual sources due to the different sampling and signal-to-noise properties of each object \citep{Anderson2014a}.
Therefore, we apply the ``bundle'' approach by \citet{Anderson2014a} which uses the flat-fielded \texttt{flt} images derived from each of the 6 exposures. Accurate (4 times supersampled) ``effective'' PSFs were constructed for the \texttt{flt} domain and are available at the Space Telescope Science Institute's WFC3 webpages\footnote{\url{http://www.stsci.edu/hst/wfc3/documents/handbooks/currentIHB/c07\_ir07.html\# 447580}} \citep{Anderson2016psf}. We chose 12 apparent isolated point sources as reference sources. Most of them are similarly faint as the target, but we also included a few moderately bright objects not afflicted by diffraction spikes. Using a pixel raster of $41\times 41$, we derived bundles for the target as well as for the 12 reference sources. 
We created stack images of each source (i) after sky subtraction but without PSF subtraction, and (ii) after PSF and sky subtraction. 
Comparing these `before' and `after' images allowed us to assess the goodness of the PSF subtraction.
In general, the \texttt{flt}-based PSF subtraction works well. In the resulting images, residuals are visually negligible for objects as faint as the target. For brighter objects, the \emph{relative} residuals (with respect to unsubtracted image) are negligible as well.  We refer to the Appendix~\label{app:psf} for the study of several reference sources. There are clear extended-emission residuals in the case of the target, see Figure~\ref{fig:targetpsfsub} in which we also show the PSF-subtraction result for a similarly faint source for comparison.\\

Our results led us to conclude that the NIR source at the location of RX\,J0806.4--4123 is definitely an extended one. 
The maximum extension of the target source is reached at a position angle of $\sim 20^{\circ}$ East of North. 
Along the line of maximum extension, the angular size of the full major axis is at least $0\farcs{8}$. At $r=0\farcs{4}$ from the center, i.e., at a distance equal to the semi-major axis, the flux has decreased to 23\% of the peak flux value (for comparison, a similar flux decrease for a reference point source, e.g., s4 in Figure~\ref{fig:psfsubtract2}, results in an ``extension'' of  $0\farcs{3}$).
In the perpendicular direction, the full minor axis was measured to be $\approx 0\farcs{5}$.
If one assumes that there is indeed an additional point source, the flux of that point source represents $\sim 20$\% of the total F160W flux according to our PSF subtraction.
In order to estimate a firm upper limit of the contribution of a possible point source to the total flux,
we slightly varied the position of the subtracted point source. Using 18 positions within $r=0\farcs{15}$ of the maximum NIR flux, we estimated a standard deviation of 2.5\% in flux. In addition, we manually increased the subtracted point source flux from the fit value to larger ones until the residual ``hole'' was inconsistent with background fluctuations of the source-free region. This criterion was met when the point source constitutes 42\% of the total flux. Therefore, the estimated conservative upper limit on the contribution of a point source is 50\%.

\subsection{Gemini observation}
RX\,J0806.4--4123 was observed in the $K_s$ band with Gemini South equipped with FLAMINGOS-2 \citep{Eikenberry2012}. 
The observations were carried out from 2017 January 14 to 2017 February 6 in Fast Turnaround observing mode. The program ID was GS-2016B-FT-23.
We obtained 492, 12, and 226 dithered exposures with integration times of 10\,s, 12\,s, and 15\,s, respectively, resulting in a total exposure time of 2.3\,h.
The field of view of the Hawaii-2 array has a diameter of $6\farcm{1}$ and a pixel scale of $0\farcs{18}$ per pixel at the f$/16$ telescope focal surface. 
The airmass of the individual exposure ranged from 1.02 to 1.32, the average airmass was 1.08\\ 

For the data reduction, we used the f2 package within the Gemini IRAF Package (version 1.12) with IRAF (version 2.16). We applied the usual corrections (dark current subtraction, bad pixel mask correction, flat-fielding). 
After constructing sky frames with \texttt{nisky}, the flat-divided sky frames were subtracted from the respective flat-divided science images. Taking into account the different exposure times, we combined the data from individual nights as well as the data from all nights. Using the \emph{Gaia} data release 1 \citep{Gaia2016b,Gaia2016a} and the Graphical Astronomy and Image Analysis Tool (GAIA) of the JAC Starlink Project \citep{gaia}, we calibrated the astrometry of the reduced images to an absolute astrometric accuracy of $0\farcs{13} (1\sigma)$.\\

We selected 16 stars (magnitudes $K_s=11.92-14.62$) from the 2MASS point source catalog \citep{Cutri2003}  with quality flag AAA and no obvious nearby neighbors in the combined image to establish a photometry scale. We applied the \texttt{daophot} and \texttt{photcal} IRAF packages and used an aperture radius of 7\,pix for the 2MASS stars (the FWHM is about 3\,pix). We determined a zero point of $25.50 \pm 0.02$. With this calibration, all but the three brightest sources (brighter than 12.35) have measured flux values within $2.2\sigma$ of their 2MASS magnitudes. The reason for the deviation of the brightest reference source is likely a slightly nonlinear behavior of the detector in some of the longer exposures due to imperfect screening. Based on the other 13 calibration stars, reliable $K_s$ magnitudes are 13.0 and fainter.\\   

\begin{figure}[t]
\includegraphics[width=8.5cm]{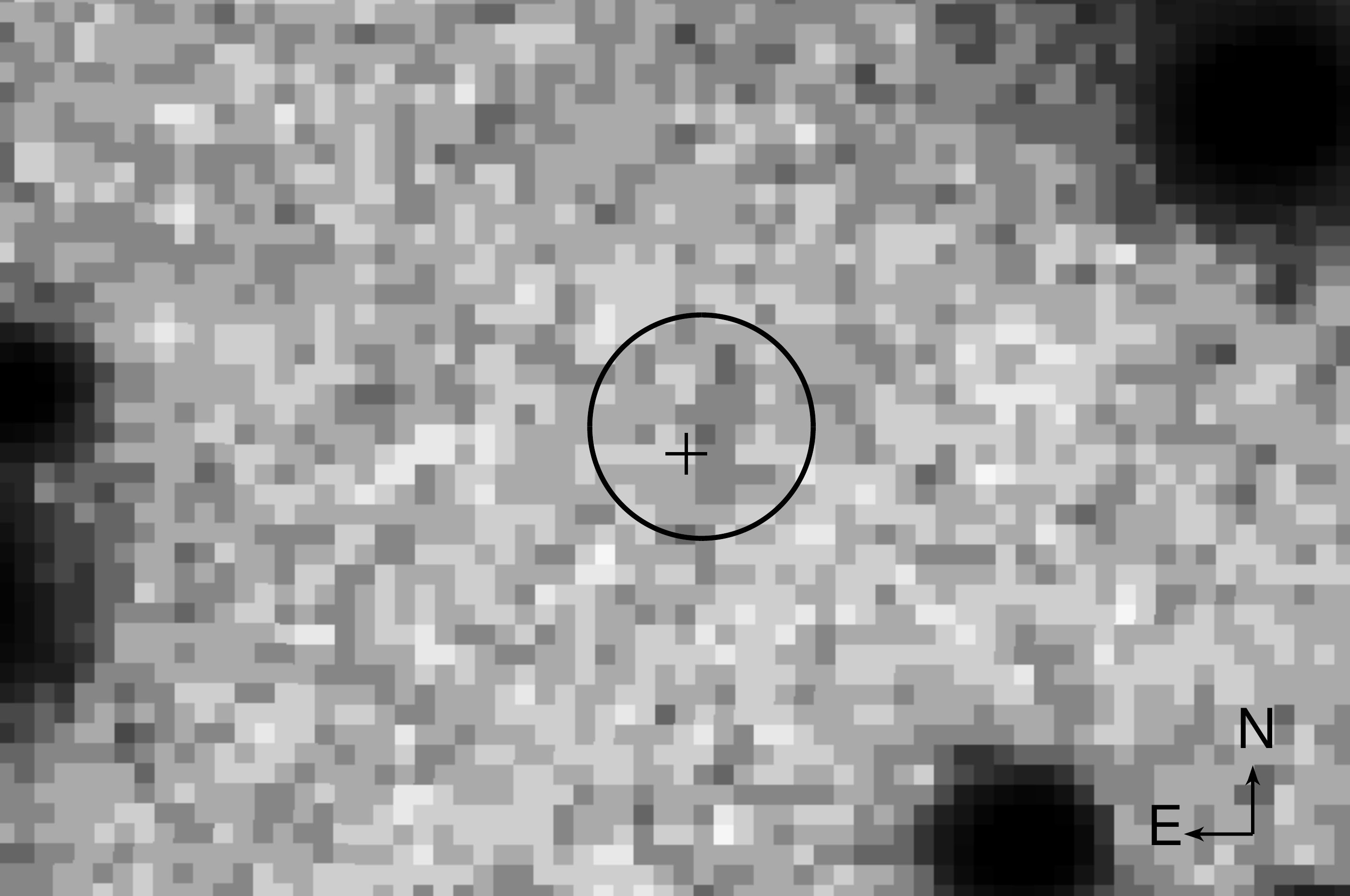}
\caption{The 2017 $K_s$ image ($8\arcsec \times 12\arcsec$), obtained with Gemini South equipped with FLAMINGOS-2.  The circle has a radius of $1\arcsec$ and marks the F160W source at the position of RX\,J0806.4--4123, and the cross marks the 2002 {\sl Chandra} position as in Figure~\ref{CXOpos}. For the Gemini data, the $3\sigma$ upper limit at that position is $K_s< 21.2$.
\label{GeminiK}}
\end{figure}
There is no significant $K_s$ source at the position of RX\,J0806.4--4123, though there may be a very faint (less than $1\sigma$) flux enhancement (see Figure~\ref{GeminiK}). 
Formally, we measured the flux in an aperture with radius of 7\,pix and obtained a magnitude $22.4 \pm 0.4$ ($1\sigma$), resulting in an upper $3\sigma$ limit of  $K_s<21.2$ (corresponding to $<2.2\mu$Jy) at the position of the neutron star. Such an upper limit is consistent with $\sim 3 \sigma$ detections in the surrounding area (radius $r \sim 30\arcsec$), where such faint sources are found to have magnitudes of 20.8 to 21.4, depending on the varying background emission.

\section{Discussion}
The NIR source at the location of RX\,J0806.4--4123 is unambiguously confirmed with our \emph{HST} F160W observations. 
The measured F160W flux, $0.40 \pm 0.06$\,$\mu$Jy, is clearly in excess of the expected neutron star flux considering the extrapolation of its UV-optical flux (see Figure~\ref{fig:sed1}).
In addition, the object has a size of at least $0\farcs{8}$ (Section~\ref{HSTnir}) and an inhomogeneous brightness distribution, consisting of a brighter `core' and elongated `halo' emission (Figure~\ref{fig:targetpsfsub}). 
In order to estimate the likelihood of an unassociated background source, we measure a source density of 0.07 sources per square arcsec 
in the F160W image which includes all sources with STmag$\leq 27.4$ (conservative limit since the target has STmag$= 27.1 \pm 0.2$). The separation of the flux peak of the extended emission from the neutron star position in the 2010 \emph{HST} F475W image is $0\farcs{15}$. Recent  \emph{Chandra} observations indicate a proper motion of RX\,J0806.4--4123 of $\approx 20$\,mas yr$^{-1}$ (Posselt et al., in prep.; the previous limit was $\mu < 86$\,mas\,yr$^{-1}$ $(2\sigma)$, \citealt{Motch2009}).
Considering a (conservative) total position uncertainty of $0\farcs{4}$ (including proper motion $2\sigma < 50$\,mas\,yr$^{-1}$, F475W position error $<0\farcs{1}$), and $N=1$ trials, we estimate a by-chance probability of $3.7$\% \footnote{Using the \emph{Chandra ACIS} position from 2015 results in a larger total position uncertainty because of the larger X-ray positional error.}. Therefore, we regard an unassociated background source as unlikely.\\ 

Even though the by-chance probability for an unassociated background source is low, a background galaxy might explain the detected NIR emission. The optical and UV data obtained with \emph{HST} (Figure\,2 by \citealt{Kaplan2011}) do not show extended emission, and the fluxes are consistent with those measured for the other six of the Magnificent Seven, i.e., it is reasonable to assume that there is no significant flux contribution from a galaxy to the F475W and L140P flux measurements of RX\,J0806.4$-$4123.
The colors of galaxies indicate that any putative galaxy counterpart must be either very red \citep[e.g.][]{Daddi2000, Franx2003, vanDokkum2006, Guo2012}, or, considering the lack of F475W flux, be at redshift $z \gtrsim 3.5$ \citep[e.g.][]{Giavalisco2002}.
The same result follows from simulations using the mock galaxy catalog by \citet{Williams2018arXiv}.
These simulations also show that the probability to detect galaxies with consistently red observed colors (either star-forming or non-starforming) in the field is low -- only $\sim 0.2$ per square arcmin are expected to have F160W fluxes within the measured flux uncertainties while also being sufficiently faint in F475W.
The analysis of actual, sufficiently deep observations results in similar numbers. Analysis of the data on the \emph{Hubble} Ultra Deep Field (HUDF; \citealt{Rafelski2015}) yield $0.39 +/- 0.17$ galaxies per square arcmin if similarly red and faint sources as our target are considered. The uncertainty results from the large Poisson error in the very small $12.8$\,arcmin$^2$ field of view.\footnote{The GOODS (e.g., \citealt{Skelton2014}) fields are larger, but the respective F475W observations are shallower than those of RX\,J0806.4$-$4123.}
Furthermore, we note that high-redshift galaxies are in general more compact, having smaller sizes than low-redshift galaxies \citep{Shibuya2015,vanderWel2014}. The detected $\sim 1\arcsec$ NIR extension would indicate an unusually large high-redshift galaxy. In the mock galaxy simulations, for example, $<5$\% of galaxies with consistent observed colors are large enough to also be consistent with the detected NIR extension.
Thus, we conclude that a high-$z$ galaxy is an unlikely counterpart of the extended NIR emission at the position of RX\,J0806.4$-$4123.\\
The detected extended emission could also be due to coincidence with emission of the interstellar medium (ISM). Scattered stellar radiation, hot dust, or line emission from the ISM gas are potential origins. We discuss these possibilities in more detail in Section~\ref{ISMdis}.\\

If the NIR emission is associated with\\ RX\,J0806.4$-$4123, then $L_{\rm F160W} \sim 6 \times 10^{27}$\,erg\,s$^{-1}$, assuming a distance of 250\,pc.
Using the X-ray flux and spin-down power of the NS, the NIR efficiency is ${\eta}_{\rm F160W} = L_{\rm F160W} /\dot{E} = 4 \times 10^{-3}$, and the luminosity ratio $L_{\rm F160W}/L_{X}= 3 \times 10^{-4}$. 
The F160W flux value, $0.40 \pm 0.06$\,$\mu$Jy, and a size of $0\farcs{5} \times 0\farcs{8}$ translate into an average brightness temperature\footnote{The brightness temperature $T_B$ is defined from the equation $F_{\nu} {\Omega}^{-1} = B_{\nu}(T_B)$, where $F_{\nu}$ is the flux density measured in the solid angle $\Omega$ subtended by the source, and $B_{\nu}(T_B)$ is the Planck function.} $T_B=282$\,K. 
If our source were a blackbody with this temperature, then one would expect a very high IR flux,
e.g., $F_{\rm 4.5 \mu m}=49$\,Jy which is clearly above the current flux limits by \emph{Spitzer}. 
However, thermal emission with a temperature, $T=950$\,K and a filling factor \footnote{The filling factor is the ratio of projected emitting area and projected source area.} of $7.3 \times 10^{-11}$ would be consistent with the F160W and $J$-band flux measurements as well as the upper limits obtained for $K_S$ and $4.5 \mu m$. 
It follows from these estimates that thermal emission is, in principle, consistent with the (N)IR measurements if the \emph{emitting} area is a very small fraction of the apparent area of the detected extended source.
Thermal emission could come from warm gas or X-ray irradiation and subsequent cooling of solid bodies such as dust or a substellar companion.
At a distance of 250\,pc, an extension of $\sim 0\farcs{8}$ corresponds to $3\times 10^{15}$\,cm or 200\,AU.
This rules out a substellar companion as the sole origin of the NIR excess.
However, the existence of a warm substellar companion cannot be entirely ruled out since the conservative upper limit on the contribution of a point source to the total flux is 50\%. Such flux limit translates into a substellar companion mass limit of  $<10$ Jupiter masses (using a bolometric correction of 3.5 \citep{Golimowski2004}, the calculations presented by  \citet{Burrows2001}, assuming a distance of 250\,pc and an age of 10\,Myr). 
Warm gas and heated dust, either in the ISM or cirumstellar matter, are discussed in Sections~\ref{ISMdis} and \ref{DISKdis}.

Alternatively, the F160W measurement can also be explained by non-thermal emission.
Assuming a power law $F_\nu \propto \nu^\alpha$ for non-thermal emission, the current spectral constraints imply either a rather steep slope $\alpha<-2.5$, or an exponential cutoff between the F160W and F475W bands. Charged particles producing curvature and/or synchrotron emission would be the likely source. With respect to an association to a neutron star, the most obvious origins are magnetospheric emission (for the possible flux contribution from a point source) or a pulsar wind and the shocks it produces. We discuss such a pulsar wind nebula in Section~\ref{PWNdis}.

\subsection{Emission from the ISM ?} \label{ISMdis}
In principle, an ISM density enhancement along the line of sight could scatter NIR emission from ambient light and produce the observed extended NIR emission. 
The so-called cloudshine is a large-scale (typically, several arcminutes) phenomenon that is caused by ambient star light around large and dense molecular clouds \citep{Foster2006,Lehtinen1996}. There are no other extended NIR features and no indication of a molecular cloud (e.g., \citealt{Dobashi2005, Dame2001}) in the field of RX\,J0806.4--4123 which one would expect if the NIR emission were due to cloudshine.\\
Instead of the interstellar radiation field and a large, dense molecular cloud, one could postulate that the NIR emission could originate from local point source emission that is scattered by a small (i.e., unaccounted for) ISM enhancement at a slightly larger distance than the one of RX\,J0806.4--4123\footnote{The absorbing hydrogen column density inferred from the X-ray spectrum of RX\,J0806.4--4123 is between 
$N_H=(0.4 \pm 0.1) \times 10^{20}$\,cm$^{-2}$ \citep{Haberl2004} and $N_H=(1.7 \pm 0.2) \times 10^{20}$\,cm$^{-2}$ \citep{Kaplan2009}, depending on the spectral model. Thus, using the relation between $N_H$ and $A_{\rm V}$ by \citet{Foight2016}, the extinction of the neutron star, $A_{\rm V}<0.1$, rules out a residence in dense ISM. The color excess along the line of sight toward the neutron star is only $E(B-V)=0.14 \pm 0.08$ ($1\sigma$) even at a distance of 1.5\,kpc (well beyond the likely distance of the XTINS) according to the a recent 3D map of the ISM by \citet{Capitanio2017} (their spatial resolution along the line of sight is $\sim 200$\,pc at $1-1.5$\,kpc).}.
However, in this case the expected thermal NIR flux from the XTINS surface alone is too low to produce the large observed NIR excess by scattering. An additional NIR emitter would be required as well as an unlikely ``wall'' in the ISM.\\

If the NIR excess flux is due to genuine emission from the ISM, it could originate either from ISM gas or dust in the diffuse ISM or in an ISM density enhancement.
There are a few emission lines of ISM gas in the H and the other NIR-bands. Strongly forbidden rotational-vibrational emission lines of H$_2$ are observed, for example, in reflection nebulae or shocked regions in star forming regions and around AGB stars
(e.g., \citealt{Shull1982}).
These $H_2$ emission lines are due to collisional excitation or UV fluorescence, and in accordance with their detection in dense environments  they require high densities and significant molecular fractions \citep{Black1987,Hollenbach1989}.
The low $N_H$ value of RX\,J0806.4--4123 implies not only a low density, but also a very low molecular fraction, $2N_{H_2} (2N_{H_2} + N_{HI})^{-1} < 10^{-4}$,
because inefficient self-shielding from cosmic rays leads to dissociation of $H_2$ \citep{Burgh2007,Savage1977}.
Therefore, the NIR excess flux around the neutron star is unlikely to be produced by H$_2$ emission lines.\\ 

Interstellar dust emission peaks in the far infrared. 
It can also emit NIR photons if it is hot enough. The required high temperatures (1000\,$-$2000\,K) in thermal equilibrium could only be reached very close to the XTINS (unresolved by \emph{HST}) by typical (size $\sim 0.1\mu$m) dust grains. A non-equilibrium process, the so-called stochastic heating of very small grains $\lesssim 10$\,$\AA$ or large polycyclic aromatic hydrogen (PAH) molecules by UV photons, was found to produce the large-scale NIR emission of reflection nebulae (e.g., \citealt{Draine2001,Sellgren1984}). 
There are two reasons why very small grains/PAH molecules of the ISM cannot be responsible for the extended emission around RX\,J0806.4--4123. First, there is not a sufficient amount of very small grains in the ISM surrounding the XTINS to produce the observed NIR flux.
Using the constraints on the XTINS spectrum in UV \citep{Kaplan2011}, and our flux measurement at $1.6\mu$m, one can estimate the required number density, $n_{\rm vsg}$, of very small grains ($\sim 10$\,$\AA$) which are stochastically heated by $\sim 10$\,eV UV photons \citep{Sellgren1984}. The value, $n_{\rm vsg} \sim 10^{-3}$\,cm$^{-3}$, is at least four orders of magnitude larger than what one can expect for the typical ISM with a hydrogen number density $n_{\rm H} \sim 1$\,cm$^{-3}$, dust-to-gas ratio of 6\%, and  typical grain size and composition distribution
(all from, e.g., \citealt{Zubko2004}; we used their BARE-GR-S dust model which matches observations of the ISM in the solar neighborhood). The lack of known ``NIR halos'' around other strong UV sources such as white dwarfs supports our estimate that the diffuse ISM does not provide enough small grains for substantially enlarged NIR flux close to these objects. 
The second reason why very small ISM grains are an unlikely origin for the observed NIR emission is the strong X-ray emission of the XTINS. 
\citet{Voit1991} estimated that X-ray photons efficiently evaporate grains of sizes $\lesssim 10\AA$. Overall, we conclude that the observed NIR emission around RX\,J0806.4--4123 is unlikely to come from either gas or dust of the photon-heated ISM. We discuss the interaction of potential pulsar wind particles with the ISM in Section~\ref{PWNdis}.

\subsection{Emission from circumstellar material?} \label{DISKdis}
If the NIR emission comes from the immediate vicinity of RX\,J0806.4--4123, it could be due to a surrounding disk or torus structure, similar to what has been inferred for the magnetar 4U 0142+61 from the \emph{Spitzer} detection by \citet{Wang2006} or for white dwarfs from NIR detections (e.g., \citealt{Melis2010}). 
The stochastic heating of very small grains, which was discussed in the previous section for the ISM, is a possibility for circumstellar matter, too. However, close to a neutron star, one expects quick destruction (evaporation by X-rays) and removal (by Poynting-Robertson drag; e.g., Figure 10 by \citealt{Posselt2014}) of such small grains. Hence, regular replenishment, e.g., by collisions of asteroids or large grains, would be required to keep such a `debris disk' bright in the NIR.\\

Considering alternatively an active fallback disk model,
\citet{Ertan2017, Ertan2014} showed that such a disk is able to produce the X-ray properties and the optical excess of the XTINSs. 
Here, we use their model to check whether such a fallback disk can reproduce the measured NIR excess. 
The model employs dissipative heating which is most relevant in the active inner part of the disk, for RX\,J0806.4--4123, from the inner disk radius, $\rin \sim 10^9$\,cm to $\lesssim 5 \times 10^{11}$\,cm. The model also considers X-ray irradiation which is most relevant for the passive outer part of the disk. For details on the model and the model parameters for RX\,J0806.4--4123, we refer to \citet{Ertan2017, Ertan2014}.
The outer radius of the passive disk
is not well known and depends on the initial disk conditions and the X-ray luminosity in the early phase of evolution. However, the local irradiation temperature is very small for large radii -- for RX\,J0806.4--4123, it decreases to $< 30$\,K for $r > 10^{13}$ cm. 
For studies of the flux at infrared and shorter wavelengths, it is sufficient to use an outer radius $\rout \sim 10^{13}$\,cm.\\

The extension of the NIR source at the distance of RX\,J0806.4--4123 is $r \sim 10^{15}$\,cm (or $\sim 100$\,AU). This raises the question of whether there is any possibility for the disk model to explain not only the observed NIR flux, but also the observed source extension.
Drawing on the findings in protoplanetary disk studies, favorable properties of the disk and viewing geometry can perhaps provide such an explanation. 
For protoplanetary disks, it was shown that in the case of a flared disk NIR emission can  be scattered and produce extended emission (e.g., \citealt{Anthonioz2015, Mulders2013}).
First, we check whether the disk model can reproduce the observed spectral energy distribution. 
Then we outline a potential explanation for the observed source extension.\\ 

Using $\rin = 10^9$\,cm and $\rout= 10^{13}$\,cm, we obtain the model spectrum given in Figure~\ref{fig:sed1}. 
The disk model curve in Figure~\ref{fig:sed1} is within the error range of the data, illustrating that the fallback disk model can reproduce the observed NIR flux.
Within this disk model, the work done by the magnetic torque heats the inner rim of the disk which then produces the optical and UV fluxes.
In contrast, the NIR emission of the model spectrum is produced from the surface of the irradiated inner disk within a few times $10^9$\,cm of the XTINS.
One implication is that a completely passive disk could also produce the NIR emission, but not the optical-UV emission.\\

One possibility to explain the observed source extension is based on the assumption of disk flaring and scattering off the outer passive disk surface. 
The thickness of a disk can be described  with the pressure scale height, $h \simeq \cs /\OmegaK$, where $\cs$ is the sound speed and $\OmegaK$ is the local Keplerian angular velocity of the disk. 
In flared disks the flaring is described by the aspect ratio, $h/r$. The outer radii of protoplanetary disks around T Tauri stars reach several hundred AU and the $h/r$ values can approach unity (see, e.g., \citealt{Dullemond2010,Armitage2010,Chiang1997} for reviews).  
\citet{Chiang1997} (CG97) calculated self-consistently  the aspect ratio and the temperature profiles of these disks.
We follow the result of CG97 for our estimates, but adapt the model parameters (luminosity, mass) to the properties of RX\,J0806.4--4123.  
CG97 found $h/r \p r^{2/7}$ for an optically thick inner disk region. Outside this region there is a radially isothermal optically thin region along which the aspect ratio increases even faster with increasing $r$ as $h/r \p r^{1/2}$.
The disk photosphere, defined as the disk surface layer corresponding to the optical depth of unity to the irradiating flux, is estimated by CG97 to be $H \simeq 4 h$.
In the luminosity regime of RX\,J0806.4--4123, $\Lstar \sim L_X = 2 \times 10^{31}$ \ergpers,  using $h/r \propto \Lstar^{1/7} M^{-4/7}$ (CG97) 
with $M = 1.4 M_\odot$, we estimate the aspect ratio of the disk for  RX\,J0806.4--4123 as  $H/r \simeq 2.7 \times 10^{-2} \rau^{2/7}$ where $\rau$ is the radial distance in AU. For $\rau = 100$, we find $H/r \simeq 0.1$.
This could be taken as a lower limit, since the outermost part of the passive disk could be optically thin to its own emission, along which $H$ increases faster with increasing $r$, as $r^{1/2}$.  
In CG97, the radially isothermal disk region has a temperature $\sim 20$\,K. For RXJ0806.4--4123, such a temperature is reached at $r \sim 8 \times 10^{12}$\,cm.
Thus, a large fraction of the outer passive disk of  RXJ0806.4--4123 could be radially isothermal with $H/r \p r^{1/2}$. 
In this case, the result is $H/r \approx 0.3$ at 100 AU. 
Hence, depending on the extension of the radially isothermal region, $H/r$ could be between 0.1 and 0.3 at the outermost disk. For our estimates  below we take $H/r = 0.2$ at $\rout \simeq 100$ AU.\\ 

The NIR luminosity illuminating and being scattered off the outer disk surface, $\Lscat $,  is a small fraction of the total NIR luminosity emitted from the inner disk, $\Lin$.  As seen from the inner disk, the outer passive disk subtends a solid angle $\Omega_\m{d}$ from the mid-plane of the disk ($\theta = 90^{\circ}$) to the surface of the outer disk ($\theta = \theta_\m{d}$). For $H/r \simeq 0.2$ at $r \simeq 100$ AU, this angle is $\theta_\m{d} \simeq 78^{\circ}$.  
Assuming an albedo of $\beta \simeq 0.9$,  
we estimate that up to 4\,\%  of the total NIR emission from the inner disk could  illuminate and be scattered by the outer disk surface, $\Lscat / \Lin \lesssim 0.04$.

The \emph{observed} ratio of the NIR flux scattered off the outer disk to the NIR flux from the inner disk ($r \sim 10^9$ cm) depends on the details of the viewing and the scattering geometry, actual $h/r$ profile, temperature and the extension of the currently passive disk.
For instance, the scattering geometry matters because the scattered emission is unlikely to be purely isotropic. The angular distribution of photons scattered
by a dust grain depends on the grain composition and grain size \citep{Mulders2013,Draine2003b}. Large grains in particular show very anisotropic scattering, where one scattering produces a conical beam with the small solid angle $\Omega_{sc}$.
The NIR light emitted from the inner disk is estimated to be scattered from a thin layer of the disk surface.
In order to estimate the contribution of the scattered component to the observed NIR flux, we use a simplified  picture and represent the total scattered emission by two narrow fan beams opening from the two surfaces of the disk.
In the case of single scatterings, the solid angle of one fan beam can be estimated by integration over the  azimuthal angle as $\Omega_\m{beam} \approx 2 \pi^{1/2} \Omega^{1/2}_{sc}$.
Inside this solid angle, the ratio of the observed scattered flux to the observed inner disk flux, $\Fscat /\Fin$, is related to the ratio of the luminosities as              
\be
\frac{\Fscat}{\Fin}  \simeq   \frac{(2 \pi /\Omega_\m{beam})}{\cos i} \frac{\Lscat}{\Lin} ,   \label{equ6}
\ee  
where $i$ is the inclination angle between the line of sight and the normal to the mid-plane of the disk. 
If we assume, for example, $\Omega_\m{beam} \sim 1$ sr, and $\cos i = 0.25$ (i.e., $i \simeq 75^\circ$), we obtain $\Fscat / \Fin \approx 1$ and  $\Fin /\Ftot \approx 0.5$ which are consistent with the observation. For this scenario, half of the H-band emission is produced at the (unresolved) inner disk, while the remaining half is the contribution from the scattered light. 
In this example, the inclination of the observer and the inclination of the scattering disk surface are very similar, and the observer remains inside the beaming solid angle of the near-side disk surface.
Instead of the simple fan-beam geometry for the scattered NIR radiation, the NIR light could actually be scattered in all directions.
Our explanation remains valid 
if the scattered flux density is still concentrated near the disk surface.
If the disk is observed at small inclination angles (close to face-on view), the contribution of the scattered light to the observed flux decreases depending on the efficiency of the ``beaming", while the flux from the inner disk increases.
In other words, the observed manifestations
of any large disks around XTINSs could vary significantly depending on the optical properties of the disks, the inclination $\cos  i$ and the observed wavelength.\\

Equation\,\ref{equ6} shows that for an observer inside the solid angle of the beamed scattering, $\Omega_\m{beam}$, the observed $\Fscat  / \Fin $ could be significantly greater than $\Lscat /\Lin$. 
In our simplistic model of an assumed scattering anisotropy,
the observer sees not the entire disk but only a fraction of the disk surface close to the line of sight, which, at a given outer radius, reduces the size of the observed extended emission in comparison to the actual size of the disk. 
However, the observed size increases with increasing contribution from multiple scatterings. 
A given light ray 
can go through multiple scatterings, both local and at different radii, which could significantly change its azimuthal direction.
The details of the extended emission depend on the actual geometry and the scattering  properties of the disk and require detailed modeling
to quantify the ``beaming" of the NIR emission and its observable azimuthal distribution.
We also note that we extrapolated the results from protoplanetary disks to our case of a potential neutron star disk. The composition of such a disk is very likely quite different from that of a protoplanetary disk. For example, one could expect a higher metallicity and possibly a higher dust-to-gas ratio in supernova fallback material. 

In addition to thermal continuum, such a disk could also show line emission.
Modeling of different disk geometries, densities and compositions is beyond the scope of this paper, in particular because we currently have a good measurement in only one photometric band of the (N)IR.\\

Alternatively to a disk with favorable geometry, scattering of the emission from the inner disk by the dust grains in an additional outer cold dusty belt or torus structure around  RX\,J0806.4--4123 could also explain the extended NIR emission. Our previous \emph{Herschel} $160\mu$m detection for the region of RX\,J0806.4--4123 had resulted in an estimate of dust grain locations at $r_d= 2.3 \times 10^{16} a_{\mu \mathrm{m}}^{-1/2}$\,cm to $ 2.2 \times 10^{15} a_{\mu \mathrm{m}}^{-1/2}$\,cm for a reasonable dust temperature range of 10 to 22\,K and in dependence of the grain size, $a_{\mu \mathrm{m}}$, in $\mu$m \citep{Posselt2014}. We had speculated about the existence of a belt structure to explain the large ratio of the luminosity at $160\mu$m to the XTINS' X-ray luminosity (for details, see \citealt{Posselt2014}). The association of the \emph{Herschel} $160\mu$m detection with RX\,J0806.4--4123 is not as firm as the \emph{Hubble} NIR detection because of the worse spatial resolution of \emph{Herschel}. Nevertheless, it is intriguing that the spatial dimensions of the potential torus are of about the same order as the NIR extension. A dusty torus alone cannot explain the NIR emission because the scattered NIR emission from RX\,J0806.4--4123 would be expected to be much fainter.
The implied two-component structure is not uncommon for debris disks around main sequence stars (e.g., \citealt{Su2013}) and may also be a possibility around a neutron star.\\

Overall, a disk could be the source of the detected NIR flux, while the observed source extension could be due to scattering by cold dust grains located at larger separations from the neutron star, either in a disk with favorable properties or in a surrounding dusty torus.

\subsection{A Pulsar Wind Nebula?} \label{PWNdis}
\begin{figure*}[t]
\begin{center}
\includegraphics[width=12cm]{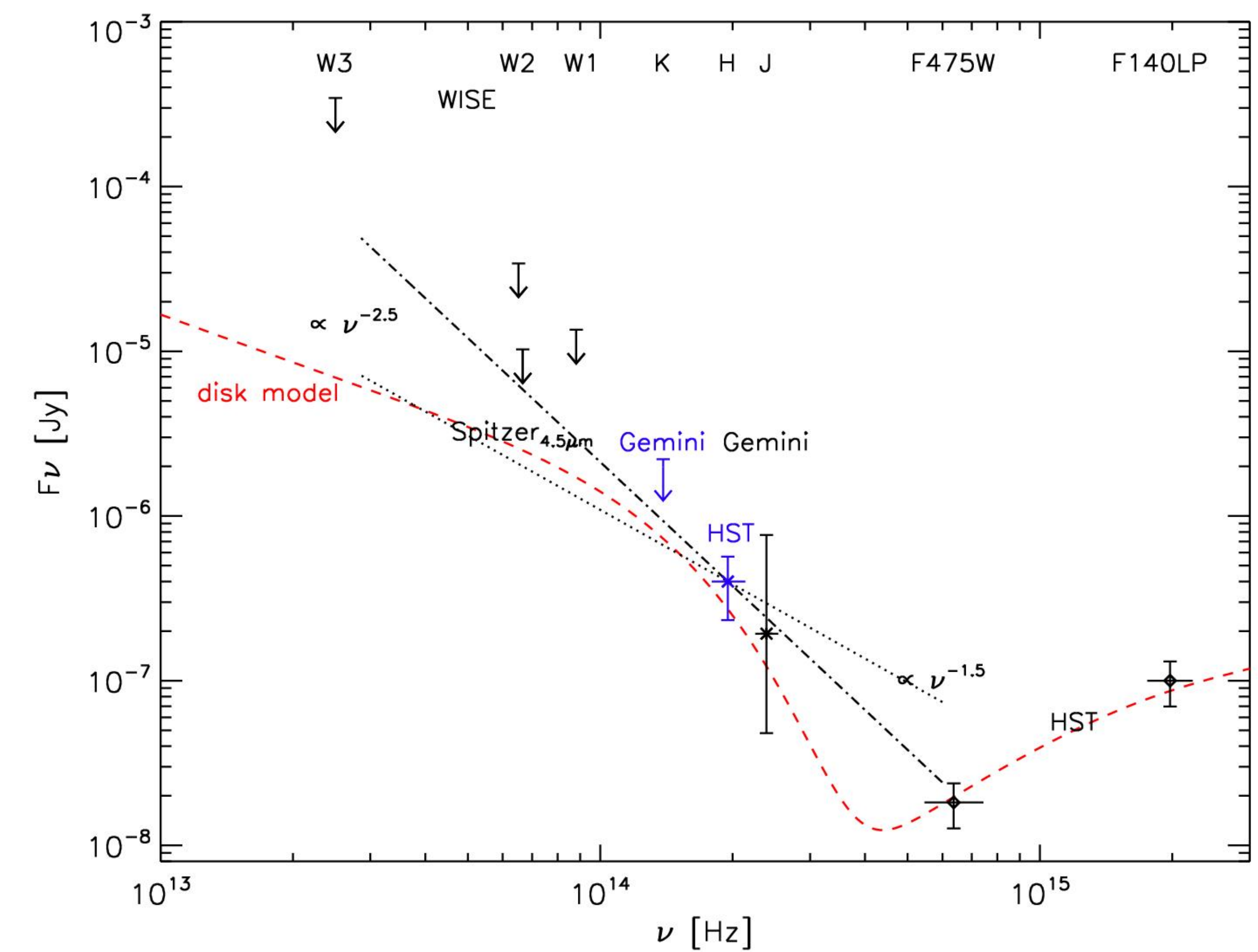}
\vspace{-0.5cm}
\end{center}
\caption{The \emph{HST} F160W flux measurement and the $3\sigma$ upper limit in the $K_S$ band from the Gemini telescope are marked in blue, the other limits and measurements are from \citet{Posselt2016}, \citet{Posselt2014}, and \citet{Kaplan2011}. The shown measurement error bars indicate $3\sigma$ uncertainties. The red dashed line shows the predicted flux of the fallback disk model discussed in Section~\ref{DISKdis}. This is the same model that was found to fit the X-ray to optical spectrum by \citet{Ertan2017} (their Figure 5). 
The plotted disk model was calculated for $\cos i = 0.5$ where $i=60\deg$ is the inclination of the observer with respect to the disk midplane normal. The NIR flux would be the same for $\cos i = 0.25$ if about half of the flux comes from scattered light from the outer disk as outlined in Section~\ref{DISKdis}.  The black dotted and dashed-dotted lines indicate different power-law spectral slopes. \label{fig:sed1}}
\end{figure*}
Extended emission around a pulsar could also be a pulsar wind nebula (PWN). 
Most PWNe are known from high-energy and radio observations (e.g., \citealt{Reynolds2017}), about a dozen have been detected in $H_{\alpha}$ \citep{Brownsberger2014}, two have recently been discovered in the ultraviolet (e.g., \citealt{Rangelov2017}).
A few high-energy PWNe were also detected in the (near)-infrared, e.g., the PWNe of Vela \citep{Shibanov2003} and PSR B0540-69 \citep{Mignani2012}, but there is only one serendipitous discovery of an infrared-only PWN with \emph{Spitzer} at wavelengths longer than $8\mu$m \citep{Wang2013}. 
The NIR emission around RX\,J0806.4--4123 could come either from line emission in the shocked ambient medium or from synchrotron emission of the shocked pulsar wind. 
In the case of shocked ambient medium, possible origins are 
emission lines of $H_2$ and $[ \rm{Fe}${\sc{ii}}$] $. These shock-excited lines are frequently observed in the dense environments of star-forming regions (e.g., Herbig-Haro objects), evolved stars with eruptive mass loss, and supernova remnants (e.g., \citealt{Lee2014,Reipurth2001}). Such a dense environment, however, is unlikely for RX\,J0806.4--4123 based on the X-ray absorption in the XTINS spectrum and 3D  models of the ISM (Sec.~\ref{ISMdis}). Hence, in the case of the PWN scenario, the emission is probably due to synchrotron emission from the shocked pulsar wind.\\

At low $\dot{E}$, particles cannot be accelerated to sufficiently high energies to produce X-ray emission, but emission at lower energies could be possible.
According to \citet{Reynolds2017}, the maximum synchrotron photon energy $E_{\rm max}$ of the accelerated electrons in the shocked pulsar wind can be estimated as 
\be
E_{\rm max} \lesssim 0.13 \zeta \dot{E}_{30} B_{-6} \frac{\sigma}{\sigma+1} {\rm eV}
\label{emax} 
\ee
where $\zeta\sim 1$ is a numerical factor, $\dot{E}_{30}$ is the spin-down power in units of $10^{30}$\,erg\,s$^{-1}$, $B_{-6}$ is the magnetic field at the location of the shocked pulsar wind in units of $\mu$G, and $\sigma$ is the magnetization which is defined as the ratio of the Poynting flux to the particle enthalpy flux. Expected PWN magnetic field values are of the order of $10-100\mu$G depending on distance, while $\sigma$ is unknown but usually assumed to be $\lesssim 1$ or even $\sigma \ll 1$ \citep{Reynolds2017,Kennel1984}. 
The spin-down power of RX\,J0806.4--4123 is $\dot{E}_{30}=1.6$ \citep{Kaplan2009}. Thus, $E_{\rm max} \sim 1$\,eV (corresponding to $\lambda \approx 1.2\mu$m) is possible for $\sigma \lesssim 1$.\\ 

The X-ray efficiencies, ${\eta}_{X} = L_X /\dot{E}$ of observed PWNe vary greatly, reaching values between $\lesssim 10^{-6}$ and $10^{-2}$ \citep{Reynolds2017}. 
The  similarly defined NIR efficiency of RX\,J0806.4--4123, ${\eta}_{\rm F160W} = 4 \times 10^{-3}$, is within the PWN efficiency range seen at other wavelengths.
For a pulsar moving supersonically with the total velocity $v_{\rm PSR}$ through the ambient medium with density ${\rho}_{\rm a}$,
the stand-off radius $R_s$ of the bow shock apex can be described by 
$R^2_s=\dot{E}_{\rm PW} f\, {(4 \pi c \, {\rho}_{\rm a} \, v^2_{\rm PSR})^{-1}}$,
where 
$\dot{E}_{\rm PW} = \xi_w \dot{E} \lesssim \dot{E}$
is the fraction of spin-down power carried away by the wind, and the factor $f$ takes into account a possible anisotropy of the pulsar wind ($f=1$ for an isotropic wind).
Using an observed source extension radius $r \sim 0\farcs{4}$ which corresponds to $R_s \sim 1.5 \times  10^{15}$\,cm (Section~\ref{HSTnir}), $\xi_w f \lesssim 1$, and ${\rho}_{\rm a}\approx n_{\rm a} m_{\rm p}$ ($m_{\rm p}$ is the mass of a proton), we obtain 
\be
n^{1/2}_{\rm a} v_{\perp} \lesssim 12 (R_s/10^{15} {\rm cm})^{-1} {\rm km\,s^{-1}\, cm^{-3/2} \,},
\label{velocity}
\ee
where $v_{\perp}$ is the transverse velocity component.
The proper motion of RX\,J0806.4--4123 is unusually low, $\mu < 86$\,mas\,yr$^{-1}$ $(2\sigma)$, with recent \emph{Chandra} observations indicating $\approx 20$\,mas yr$^{-1}$ (Posselt et al., in prep.). At a distance of 250\,pc, these values correspond to $v_{\perp}< 100$\,km\,s$^{-1}$ $(2\sigma)$ and 24\,km\,s$^{-1}$, respectively. This does not significantly exceed the speed of sound in the typical diffuse ISM ($c_s \sim 10$\,km\,s$^{-1}$ for $T \sim 8000$\,K). 
The constraint on the velocity and number density in Equation~\ref{velocity} depends on the actual $R_s$, but it is uncomfortably low, particularly considering that an anisotropic wind and a typically assumed fraction of the spin-down power emitted with the wind ($\xi_w \sim 0.5$) will lower the value further. $R_s$ could, however, be smaller since it is difficult to establish the exact location of the pulsar in the extended NIR emission. 
Hence, a PWN can currently not be ruled out as a possible origin of this emission.\\ 

If RX\,J0806.4--4123 were a usual rotation-powered pulsar, the PWN would be an obvious explanation.
Such an explanation is, however, somewhat surprising for a member of the Magnificent Seven which show only thermal X-ray and no radio or $\gamma$-ray emission.
A part of the detected NIR emission (the point-source contribution that can be at maximum 50\%) could in principle come from the magnetosphere of  RX\,J0806.4--4123.
However, the optical-UV spectrum of the neutron star, $F_{\nu} \propto \nu ^{\alpha}$ with $\alpha=1.62 \pm 0.20 (1\sigma)$ \citep{Kaplan2011}, is consistent with, and thus likely dominated by, a Rayleigh-Jeans tail of thermal emission (for which $\alpha=2$). 
If there is a PWN, then a lack of detectable non-thermal emission from the neutron star at higher energies than NIR is unprecedented. 
It is noteworthy that the XTINS spectra of other members of the Magnificent Seven, in particular RBS\,1774\footnote{Coincidentally, RBS\,1774 is the other XTINS where there is a potential \emph{Herschel} detection \citep{Posselt2014}.}, show much stronger deviation from a Rayleigh-Jeans tail, $\alpha=0.53 \pm 0.08 (1\sigma)$ \citep{Kaplan2011}. 
However, as discussed by \citet{Kaplan2011}, a magnetospheric origin of the deviations from Rayleigh-Jeans is not the only possible explanation, in particular considering the high optical luminosities of these XTINSs. 
The current uncertainties in the models of strongly magnetized neutron star atmospheres at long wavelengths could be, for example, another explanation for the unusual spectra of the Magnificent Seven.
If the extended NIR emission around RX\,J0806.4--4123 is indeed a PWN, it could indicate an interesting new avenue to study the properties of pulsar winds without very energetic particles originating from the (inner) magnetosphere. 
It is interesting that in comparison to typical spectral slopes of X-ray detected PWNe, 
$\alpha_{\rm X} \approx -0.5$ \citep{Kargaltsev2008},
the slope of the NIR PWN, $\alpha_{\rm PWN} \lesssim -2.5$, is rather steep according to Figure~\ref{fig:sed1}. Such a steep slope could be explained if the NIR photons correspond to the highest electron energies in the PWN.\\ 

\subsection{Is RX\,J0806.4--4123 special?}
RX\,J0806.4--4123 has a very low transverse velocity. Perhaps it also has a viewing geometry favorable for the detection of the fallback disk. Could these properties be the reason why we have seen extended NIR emission only for this neutron star so far?
Looking at the Magnificent Seven with comparably deep NIR limits, we compare the two brightest and best studied members with RX\,J0806.4--4123.
RX\,J1856.6--3754 at a parallactic distance of $123^{+11}_{-15}$\,pc, has a transverse velocity of $192^{+17}_{-28}$\,km\,s$^{-1}$ and $\dot{E} \sim 3 \times 10^{30}$\,erg\,s$^{-1}$, and its $H$-band limit is $21.5$\,mag \citep{Tetzlaff2011,Walter2010,Posselt2009,Kerkwijk2008}.
It also has a rather puzzling $H_{\alpha}$ nebula \citep{Brownsberger2014,Kerkwijk2008,Kerkwijk2001}.
RX\,J0720.4-3125 at a parallactic distance of $280^{+210}_{-85}$\,pc, has a transverse velocity of $143^{+108}_{-44}$\,km\,s$^{-1}$ and $\dot{E} = 4.7 \times 10^{30}$\,erg\,s$^{-1}$, and its $H$-band limit is $23.1$\,mag \citep{Tetzlaff2011,Kaplan2005,Posselt2009}.
In comparison to the Vega magnitude of $23.7\pm 0.2$ 
measured for RX\,J0806.4--4123 with WFC3/F160W, the $H$-band flux limits are shallower for RX\,J0720.4--3125 (factor 2) and RX\,J1856.6--3754 (factor 8).
The smaller distance of RX\,J1856.6--3754 is not small enough to counterbalance the effect of its shallower observation, while for RX\,J0720.4--3125 the large distance uncertainty makes any NIR luminosity limit highly uncertain.  
Thus, there is currently no indication that the NIR emission around RX\,J0806.4--4123 is exceptional among the Magnificent Seven.\\

Reports of NIR-detected PWNe (or disks) are rare for other isolated neutron stars.
Rotation-powered pulsars and magnetars are detected in the NIR, but there is neither such a steep slope between the optical and the NIR fluxes nor are the PWNe only detected at these wavelengths (e.g., \citealt{Danilenko2011,Durant2005, Durant2004}). 
Magnetars are on average at several kpc, hence any small-sized extended emission would be difficult to establish.
An example of a nearby rotation-powered pulsar is Geminga which is similarly distant  ($d=250^{+230}_{-80}$\,pc) as RX\,J0806.4--4123, but has a transverse velocity of $\approx 211$\,km\,s$^{-1}$ and a much higher spin-down power $\dot{E}=3.3 \times 10^{34}$\,erg\,s$^{-1}$ \citep{Verbiest2012,Faherty2007,Bertsch1992}. 
As  Figure 6 by \citet{Danilenko2011} shows, Geminga's optical-IR spectrum is relatively flat in $F_{\nu}$ though interestingly there may be some indication of a rise toward the mid-infrared. 
Geminga has a prominent X-ray PWN, and \citet{Shibanov2006} reported a detection of the PWN bow shock in the $I$ band, but not in the \emph{HST} NICMOS F160W image. From the higher $\dot{E}$ of Geminga, one would expect PWN emission at shorter wavelengths (equation~\ref{emax}). The situation could be different for rotation-powered pulsars with $\dot{E}$ as low as those of the Magnificent Seven.
The old PSR\,J0108$-$1431 has $\dot{E}= 5.8 \times 10^{30}$\,erg\,s$^{-1}$, only slightly larger than the value of RX\,J0806.4--4123. 
At a parallactic distance of $240^{+124}_{-61}$\,pc, its transverse velocity is $194^{+104}_{-51}$\,km\,s$^{-1}$ and its $H$-band limit is $21.4$ \citep{Deller2009,Posselt2009}. Again, the relatively shallow observations do not allow a  meaningful comparison with RX\,J0806.4--4123.

\section{Conclusions}
The extended NIR emission around the isolated neutron star RX\,J0806.4--4123 is unlikely to come from an unrelated field object. It can be interpreted as coming either from a disk with favorable viewing geometry or a PWN created by shocked pulsar wind particles of relatively low energy. A flux contribution of up to 50\% could come from a point-like source which could represent the unresolved inner disk or the pulsar magnetosphere in the respective interpretations.
The spectra and the details of the spatial shape are expected to be quite different for the two explanations and can be probed with future high-resolution observations with the \emph{James Webb Space Telescope}.\\

Most pulsars do not have deep NIR observations, mostly because the emission of the neutron stars at these wavelengths was expected to be just an unexciting extension of the optical-UV spectral slope.  
RX\,J0806.4--4123 is a good example that neutron stars keep surprising us.
Whether the detected extended emission around RX\,J0806.4--4123 is a lucky coincidence of low pulsar speed and favorable viewing geometry or this is a common property among (some types of) pulsars needs to be checked with systematic deep NIR surveys.

\acknowledgments
We thank the referee, Marten van Kerkwijk, for his comments and suggestions which helped to improve the quality of this manuscript.
We are very thankful to Jay Anderson for his helpful support in the WFC3/F160W data reduction, in particular his crucial help with respect to an optimal PSF subtraction. 
We also thank Mario Gennaro and Howard Bond for helping us to optimize our observation setup.

B.P. acknowledges support for this work (program 14745)
by NASA through a grant from the Space Telescope Science Institute, which is operated by the Association of Universities for Research in Astronomy, Inc., under NASA contract NAS 5-26555.

\"{U}.E. acknowledges research support from
T\"{U}B{\.I}TAK (The Scientific and Technological Research Council of
Turkey) through grants 117F144  and and from Sabanc\i\ University.  

C.C.W acknowledges support from the National Science Foundation Astronomy and Astrophysics Fellowship grant AST-1701546.

The Center for Exoplanets and Habitable Worlds is supported by the
Pennsylvania State University, the Eberly College of Science, and the
Pennsylvania Space Grant Consortium. 

This work has made use of data from the European Space Agency (ESA)
mission {\it Gaia} (\url{https://www.cosmos.esa.int/gaia}), processed by
the {\it Gaia} Data Processing and Analysis Consortium (DPAC,\url{https://www.cosmos.esa.int/web/gaia/dpac/consortium}). Funding
for the DPAC has been provided by national institutions, in particular
the institutions participating in the {\it Gaia} Multilateral Agreement.

\facilities{HST (WFC3), Gemini (FLAMINGOS-2)}

\software{DrizzlePac within PyRAF \citep{Gonzaga2012}, BUNDLE software \citep{Anderson2014a},
	  IRAF \citep{Tody1986},
	  GAIA of the JAC Starlink Project \citep{gaia}
          }

\bibliographystyle{aasjournal}
\bibliography{rxj0806}

\appendix
\section{PSF subtraction on the FLT images}
\label{app:psf}

\begin{figure}[h]
\includegraphics[width=15cm]{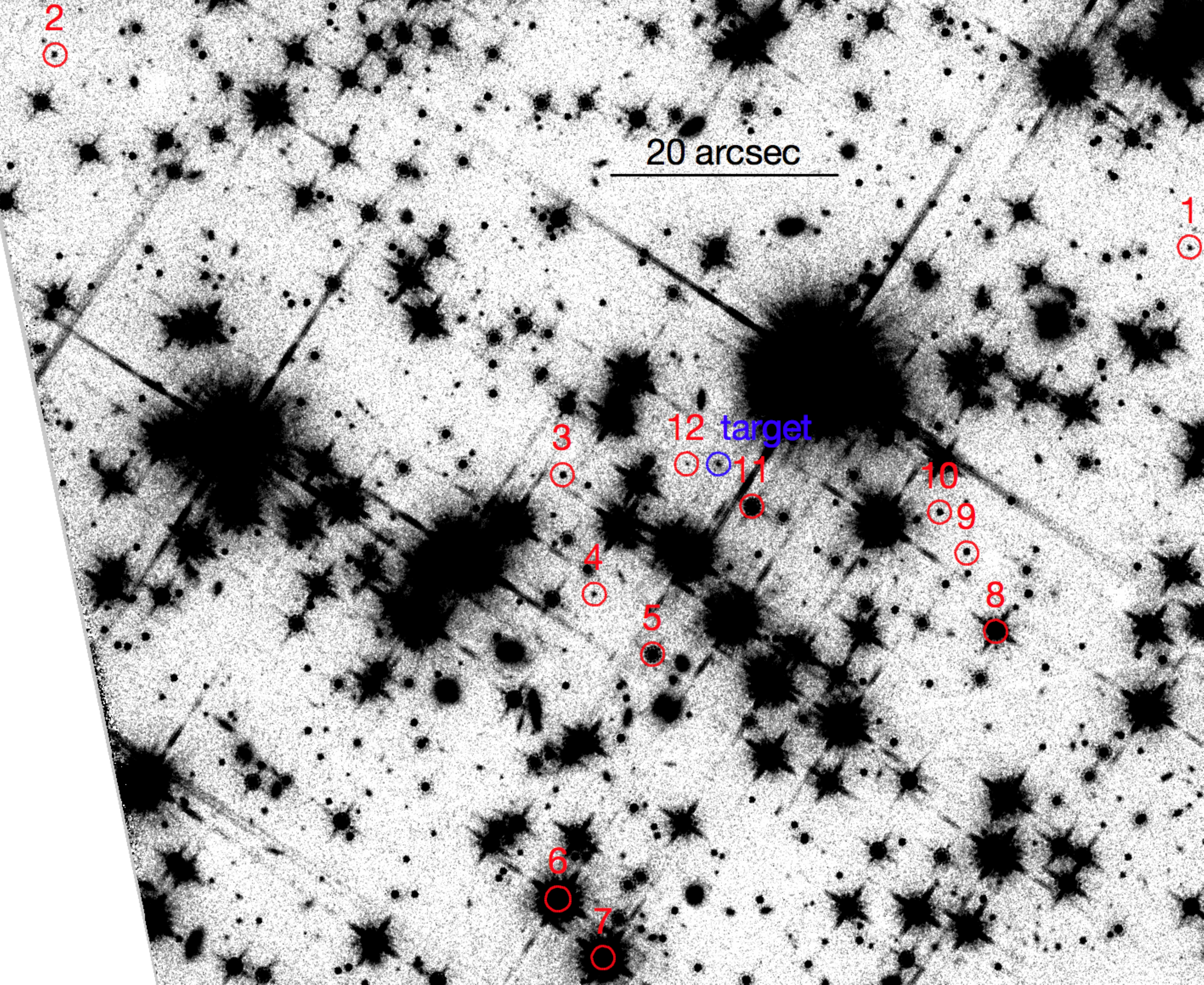}
\caption{The location of the reference sources in the WFC3/F160W observation. North is up, east to the left.
\label{whereref}}
\end{figure}

We used several reference sources to check and evaluate the performance of the PSF subtraction. Their locations in the WFC3/F160W image are indicated in Figure~\ref{whereref}.
Figures~\ref{fig:targetpsfsub}, \ref{fig:psfsubtract1} and \ref{fig:psfsubtract2} show the corresponding images before and after the PSF subtraction.
Aiming for a representation of all reference sources and the target in \emph{one} plot, we obtained image statistics in the central $0.7\arcsec \times 0.7\arcsec$ region of each bundle-stack image for both, before and after the PSF subtraction. If a source is not well subtracted one expects to see abnormally high mean and possibly abnormally high standard deviations in the sky/PSF-subtracted image in comparison to the reference sources. Since we consider a magnitude range (STmag from 19.0 to 28.7), we weigh the PSF-subtracted values with the respective ones from the image where the PSF was not subtracted. The resulting plot can be seen in Figure~\ref{fig:stat}. The measures for the target clearly deviate (larger mean, larger standard deviation) from those of the other reference sources. 

\begin{figure*}
\gridline{\fig{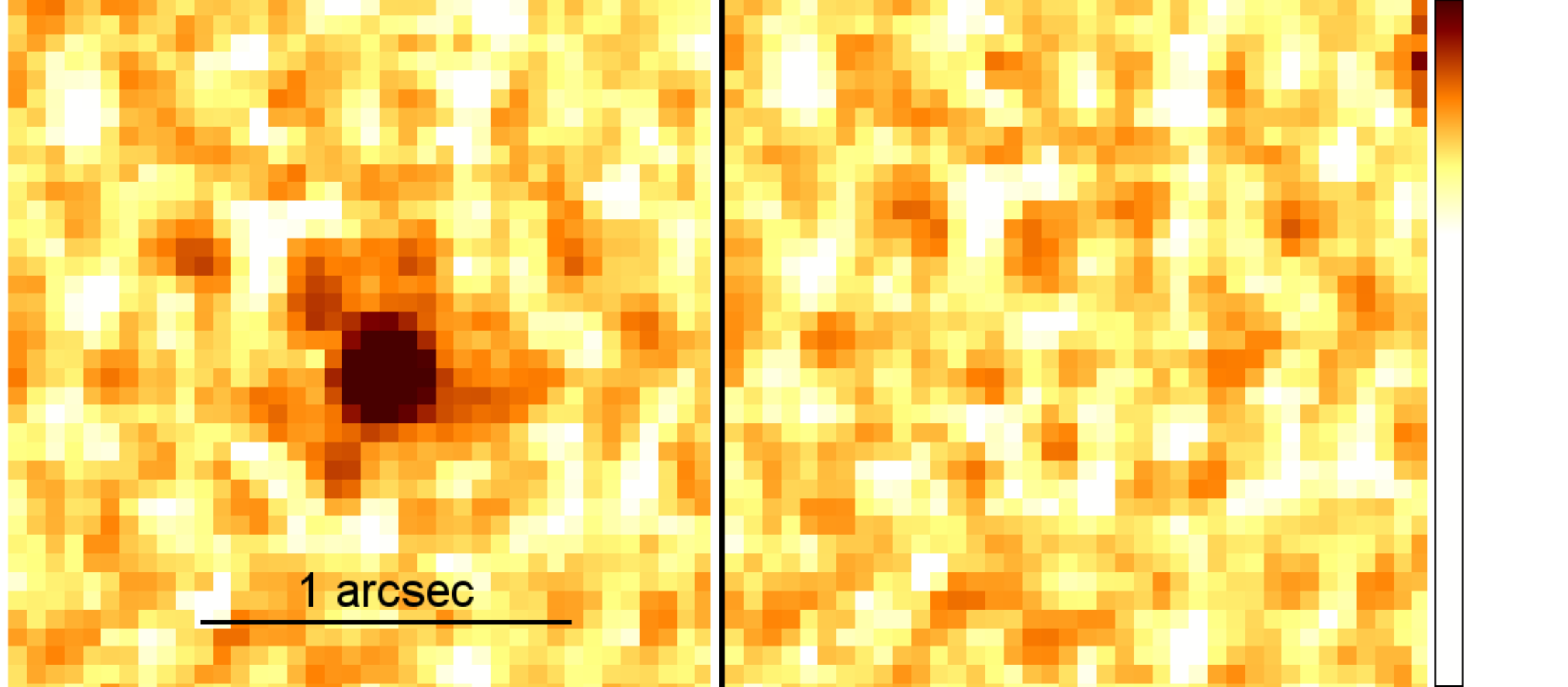}{0.49\textwidth}{reference source 1, STmag$=27.3$, R$_{\rm maxmin}=0.15$}
          \fig{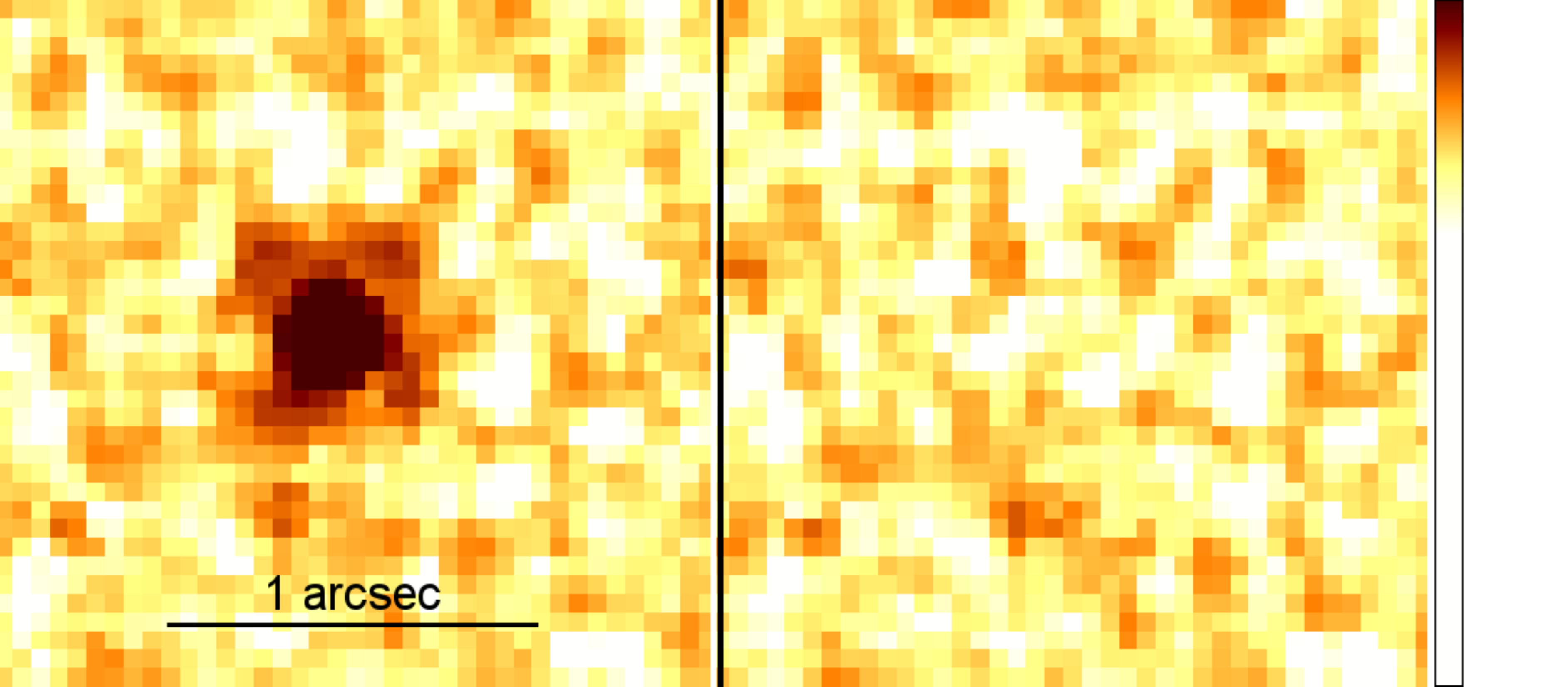}{0.49\textwidth}{reference source 2, STmag$=27.1$, R$_{\rm maxmin}=0.09$}}
\gridline{\fig{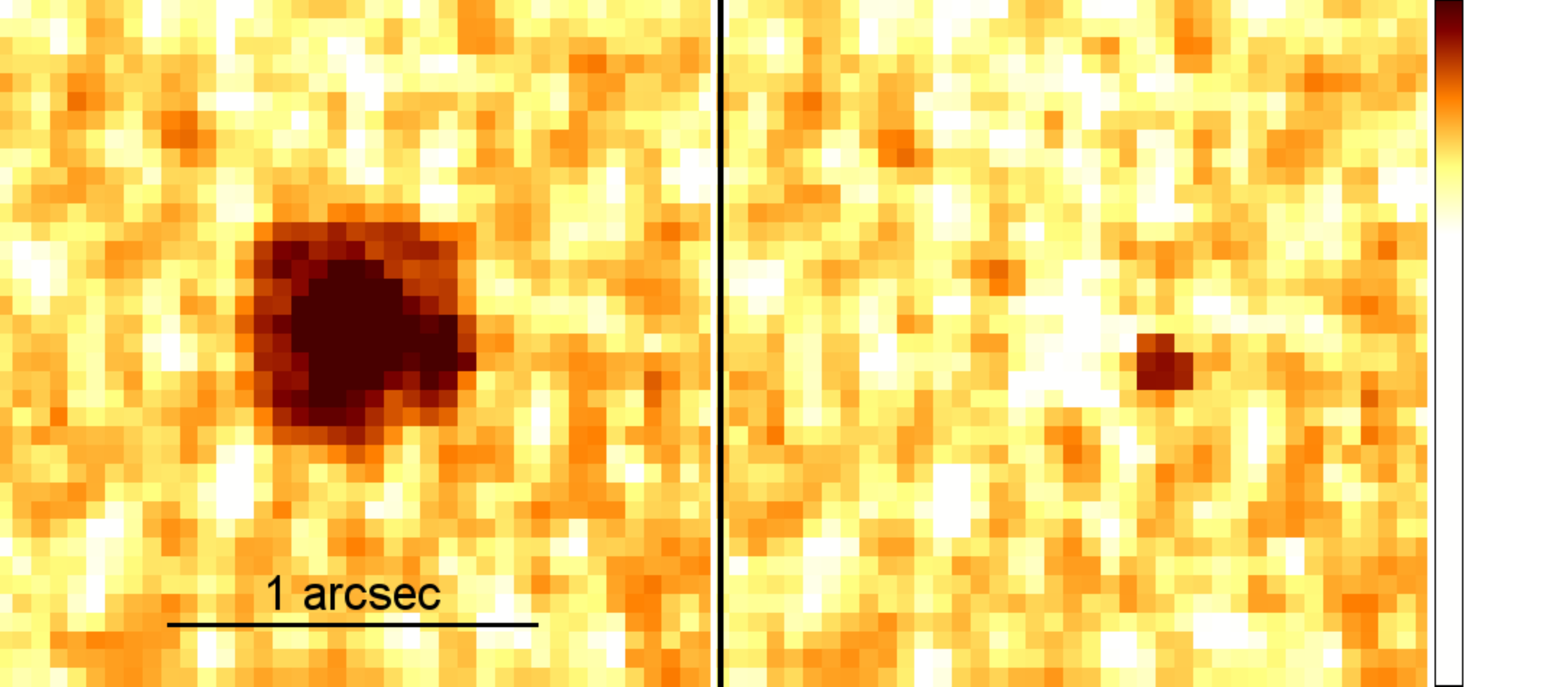}{0.49\textwidth}{reference source 3, STmag$=26.5$, R$_{\rm maxmin}=0.09$}
          \fig{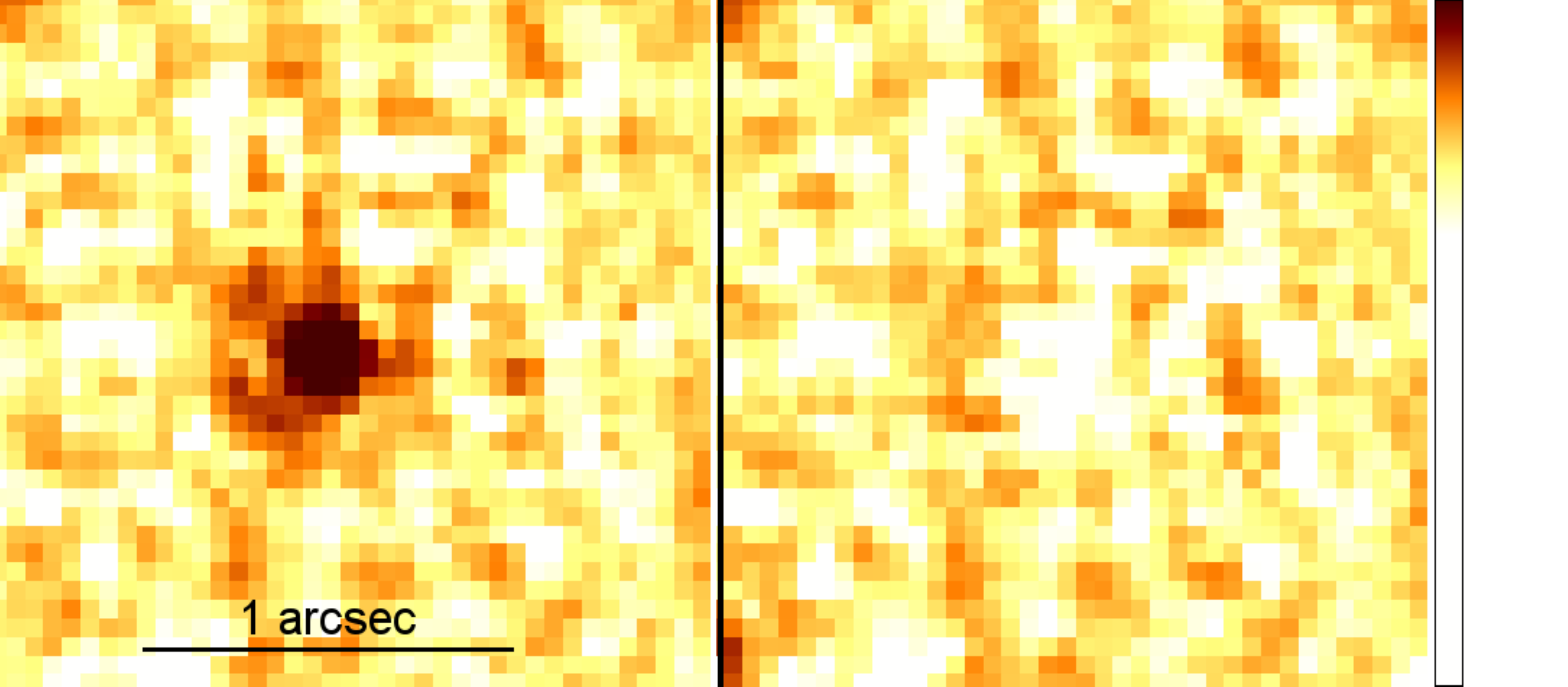}{0.49\textwidth}{reference source 4, STmag$=27.5$, R$_{\rm maxmin}=0.17$}}
\gridline{\hspace{-0.4cm}
          \fig{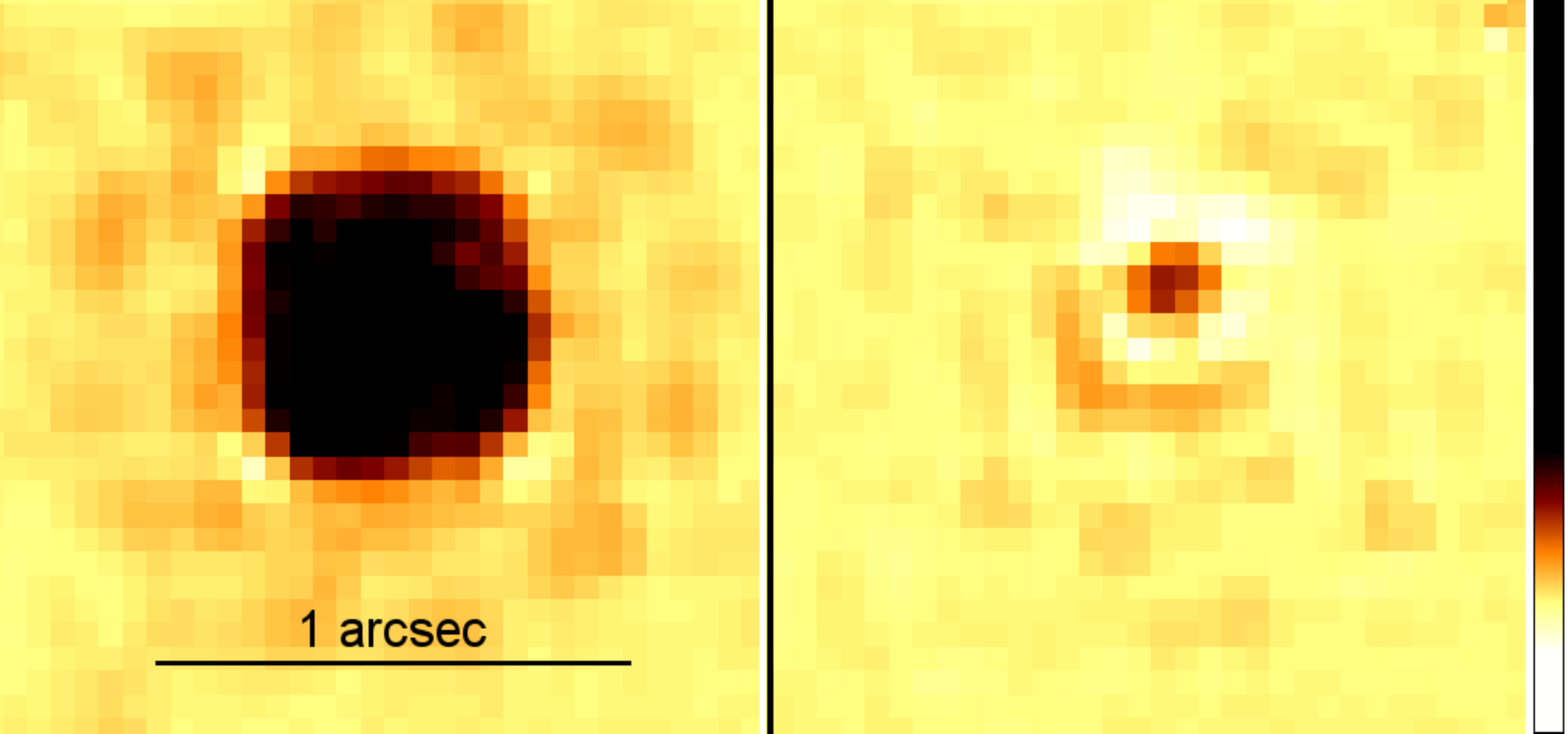}{0.46\textwidth}{reference source 5, STmag$=23.1$, R$_{\rm maxmin}=0.03$}
          \hspace{-0.4cm}
          \fig{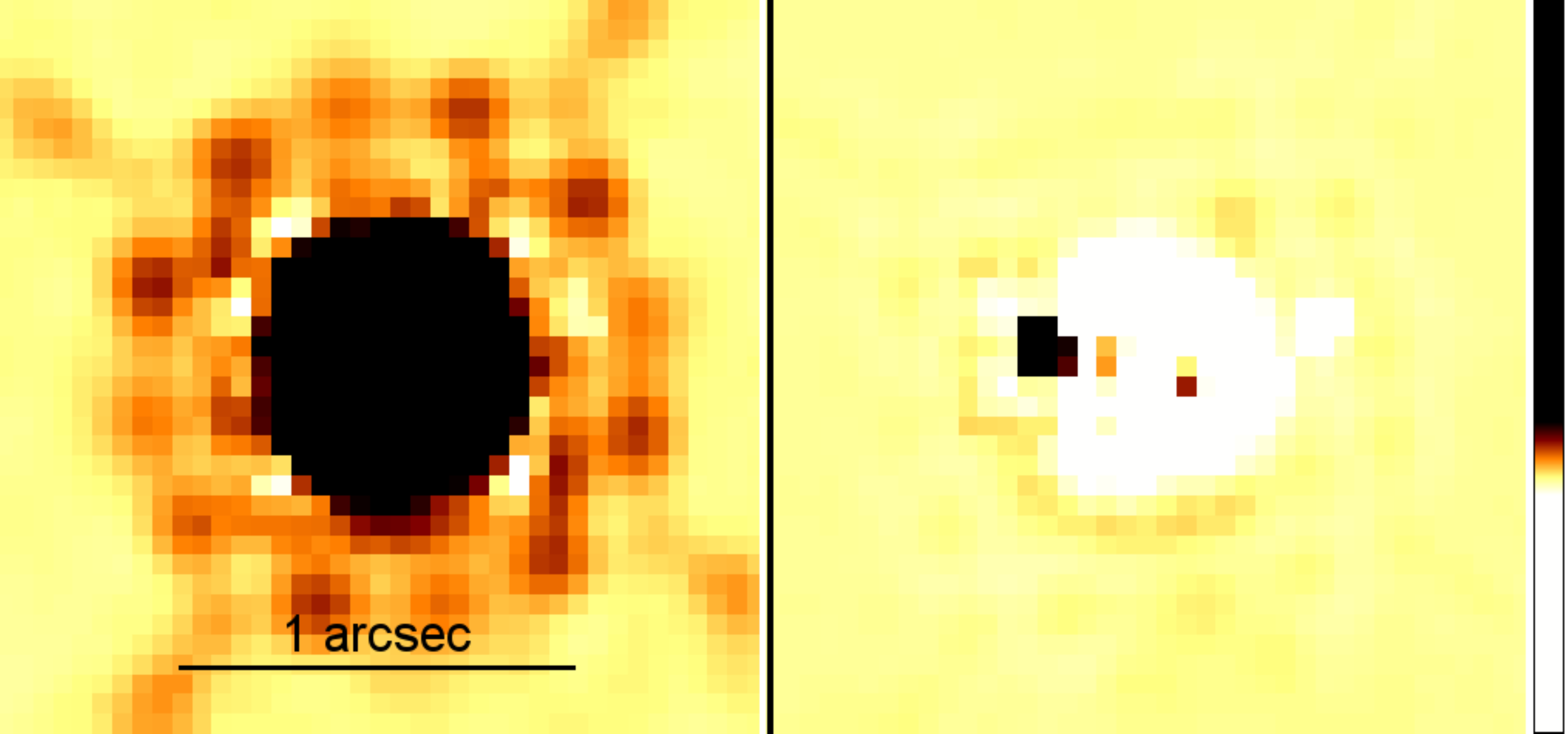}{0.46\textwidth}{reference source 6, STmag$=19.1$, R$_{\rm maxmin}=0.04$}}
\gridline{\hspace{-0.4cm}
          \fig{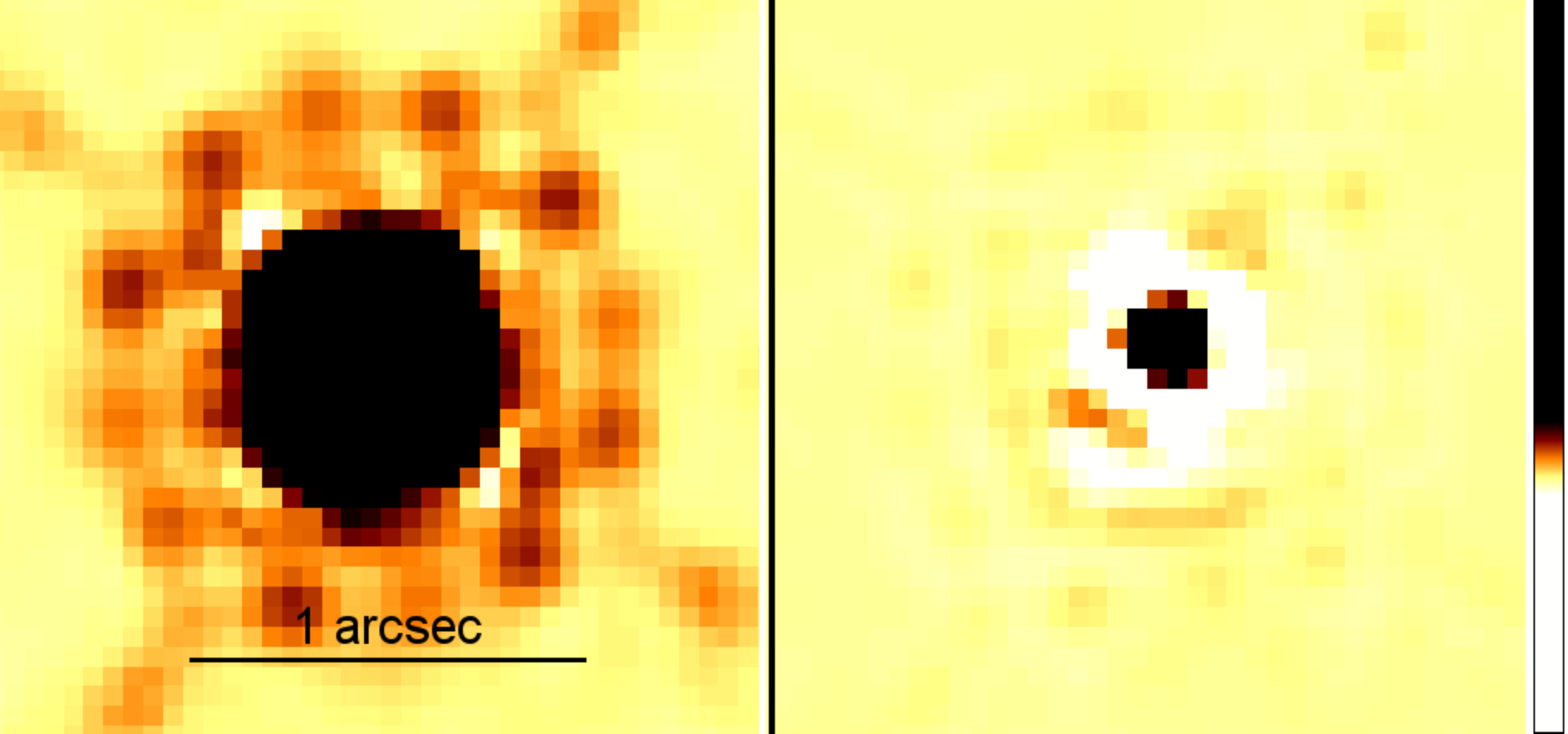}{0.46\textwidth}{reference source 7, STmag$=19.0$, R$_{\rm maxmin}=0.06$}
          \hspace{-0.4cm}
          \fig{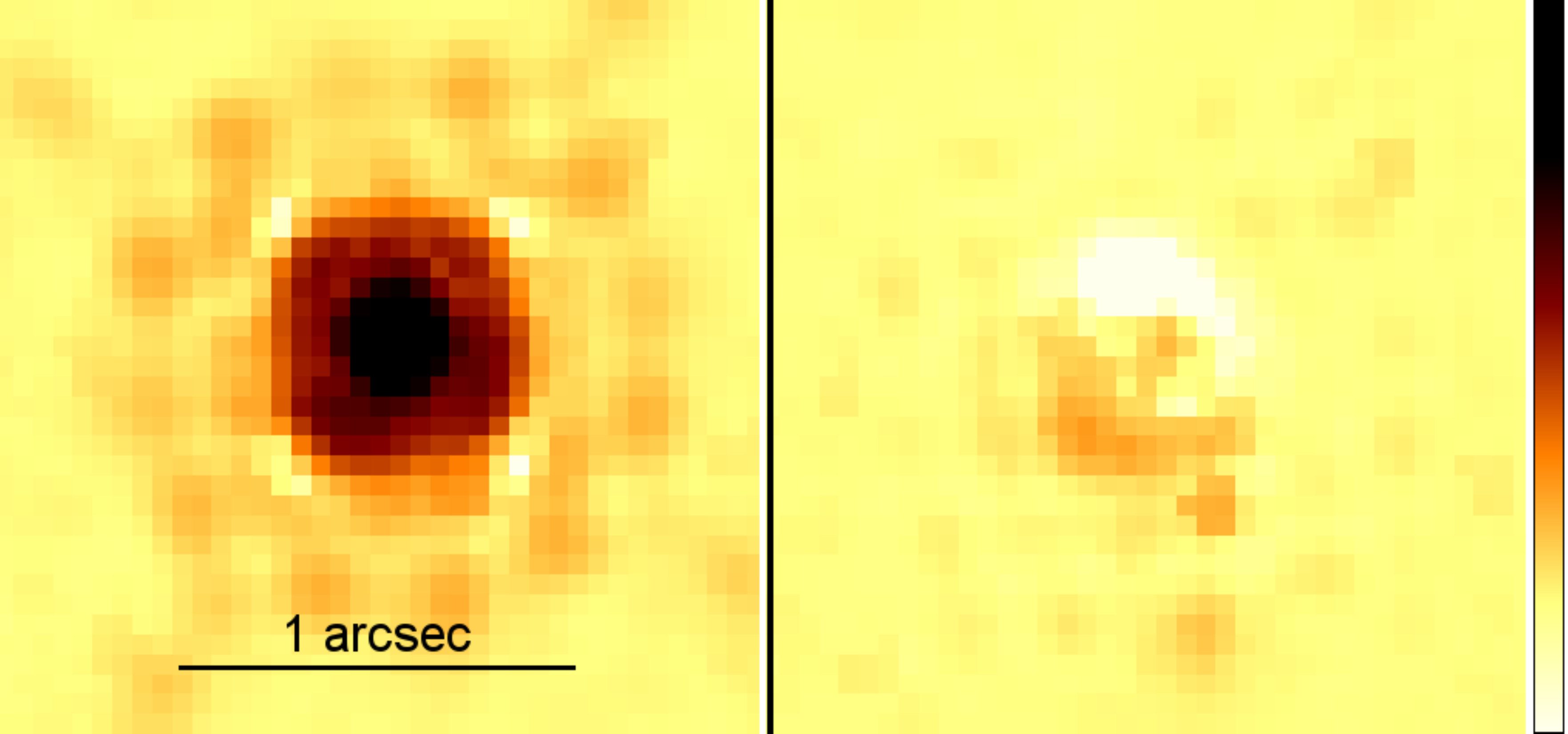}{0.46\textwidth}{reference source 8, STmag$=21.5$, R$_{\rm maxmin}=0.01$}}
\caption{The goodness of the PSF-subtraction for point sources using the Bundle software on the FLT-images. R$_{\rm maxmin}$ is defined as the ratio of the maximum$-$minimum in the PSF-subtracted image (right subpanels) to the maximum$-$minimum in the sky-subtracted image (left subpanels). The statistics was obtained in the central region using a square with $0\farcs{7}$ length. Reference sources 1 to 4 and 12 are plotted using the same limits and the same color maps as for the target. For the other reference sources, we adapted the limits and color maps to better show the larger dynamical ranges.\label{fig:psfsubtract1}}
\end{figure*}

\begin{figure*}
\gridline{\fig{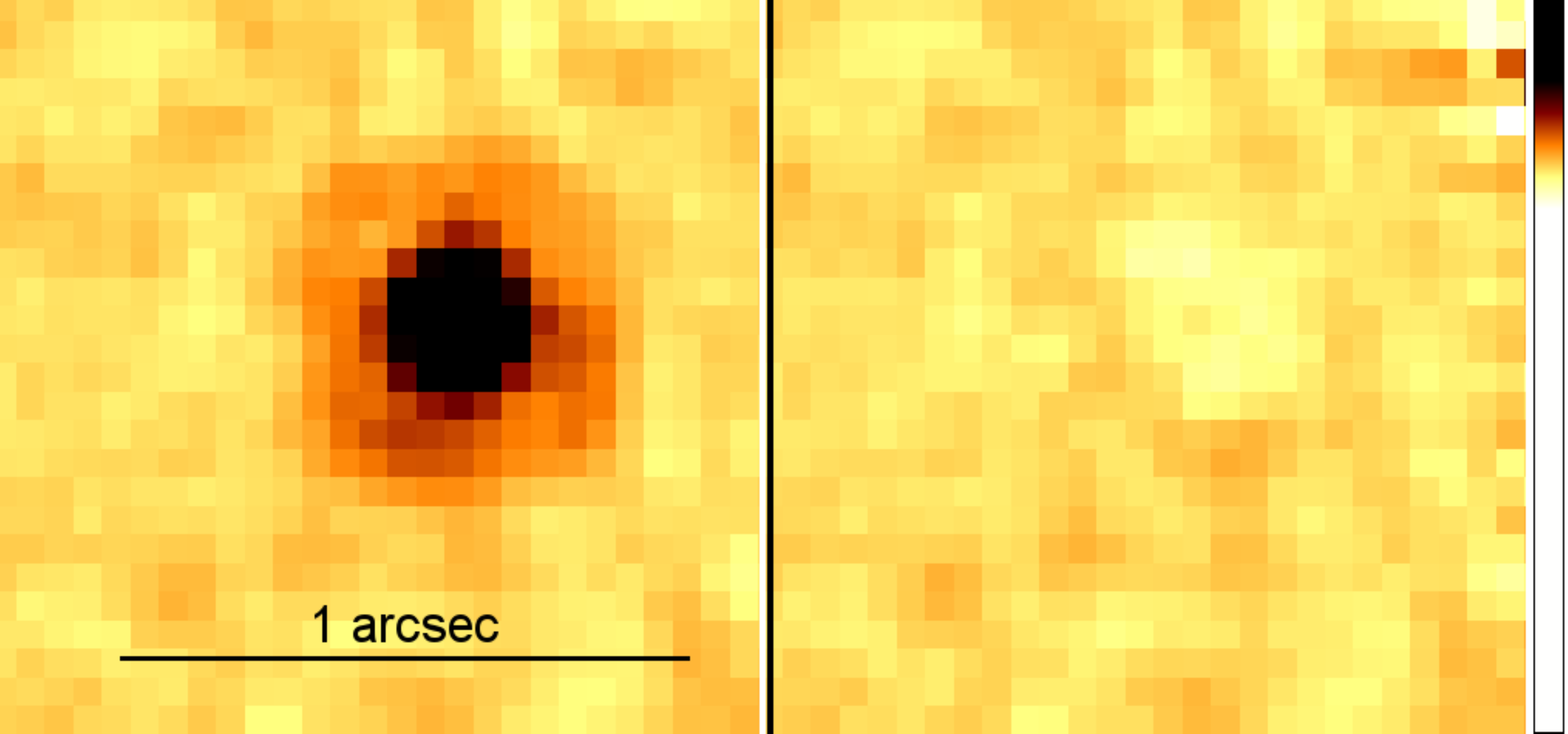}{0.49\textwidth}{reference source  9, STmag$=26.5$, R$_{\rm maxmin}=0.07$}
          \fig{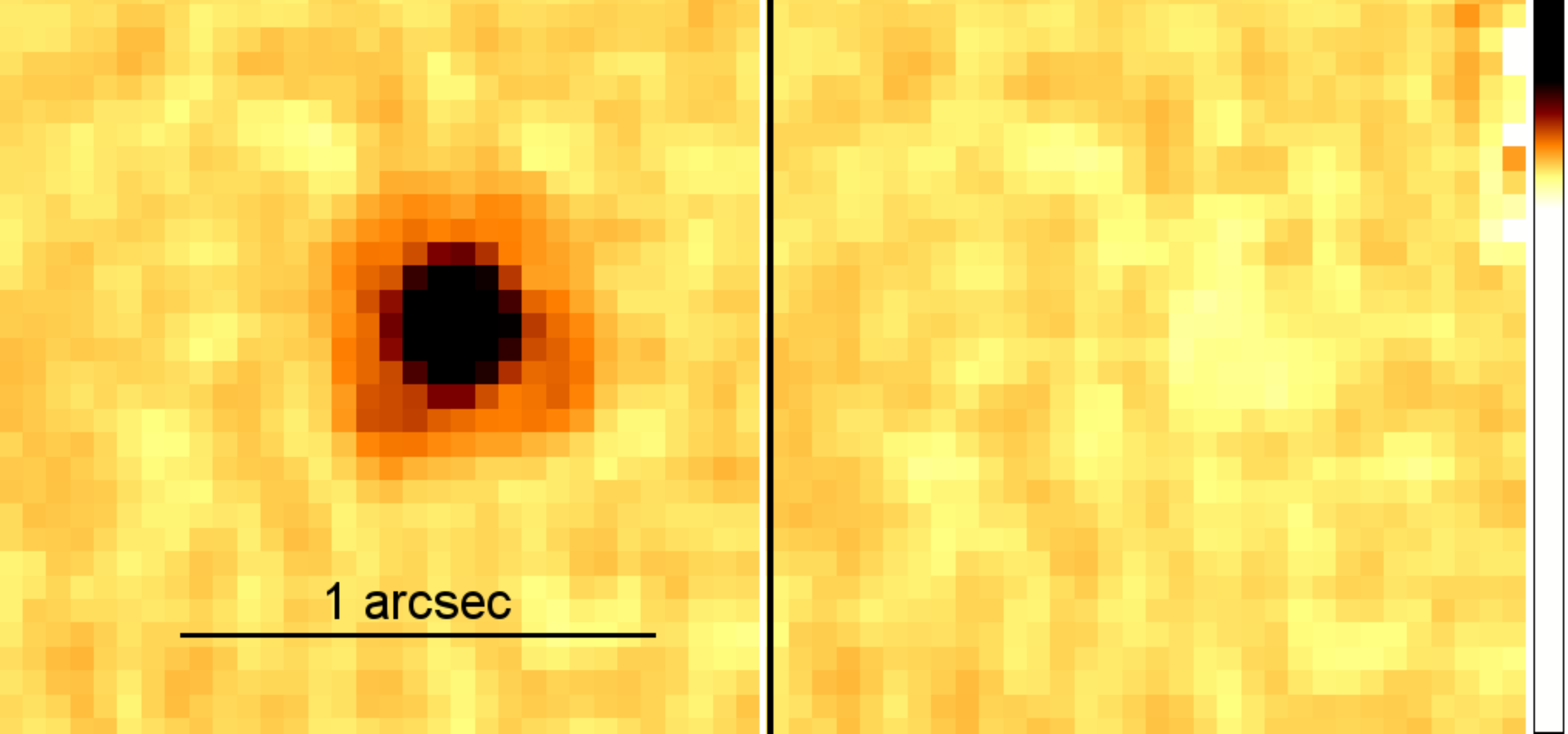}{0.49\textwidth}{reference source 10, STmag$=26.5$, R$_{\rm maxmin}=0.05$}}
\gridline{\fig{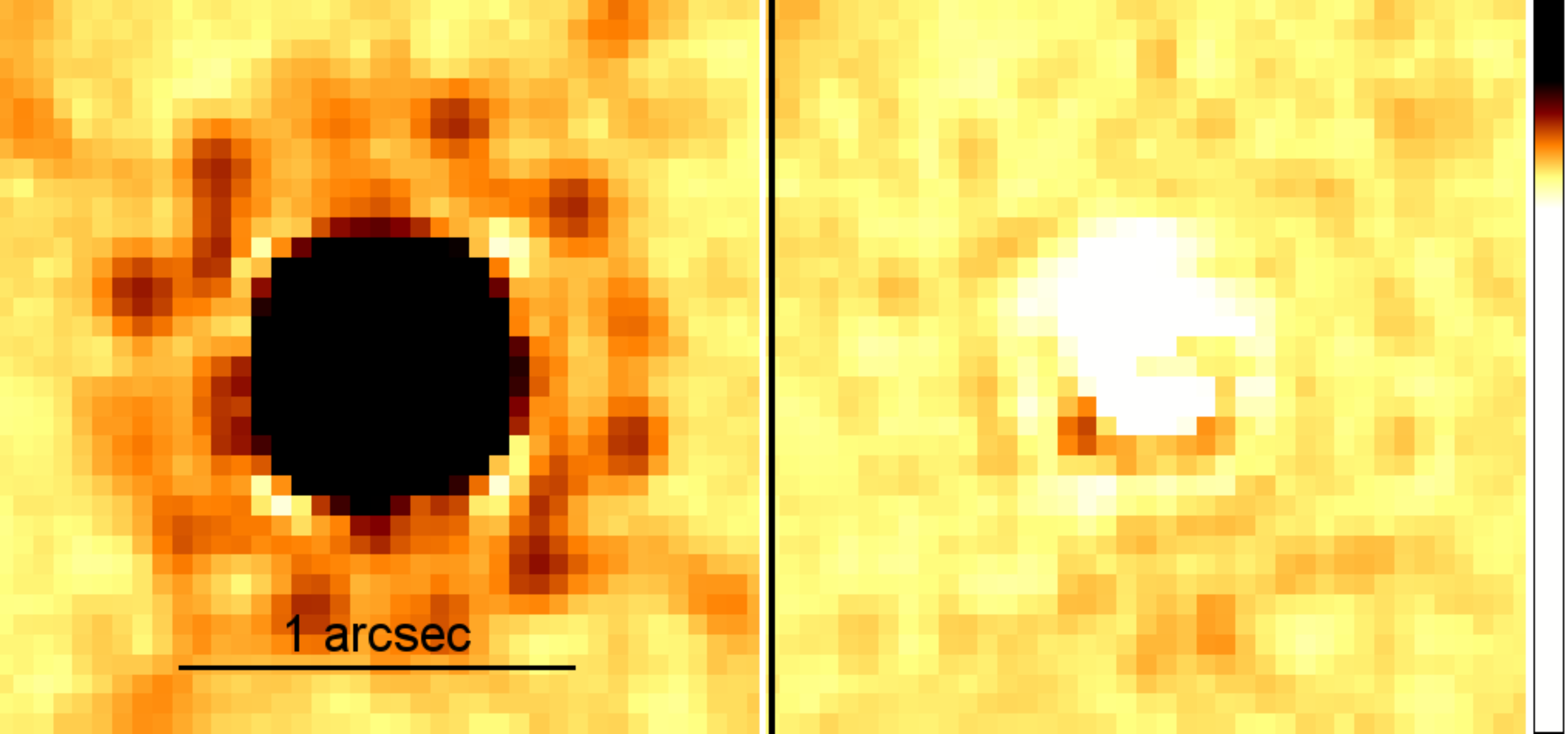}{0.49\textwidth}{reference source 11, STmag$=22.6$, R$_{\rm maxmin}=0.01$}
          \fig{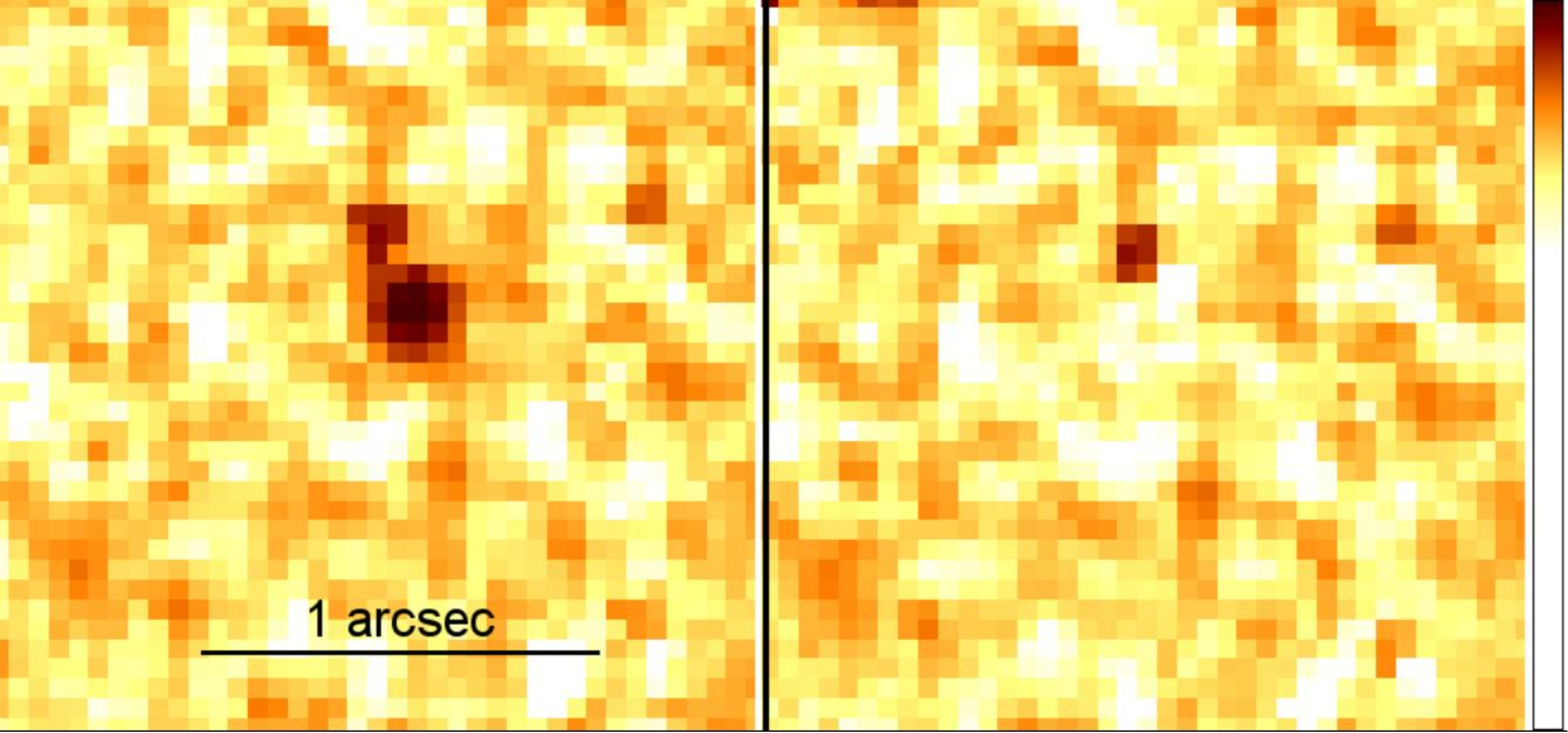}{0.49\textwidth}{reference source 12, STmag$=28.7$, R$_{\rm maxmin}=0.62$}}
\caption{Continuation of Figure~\ref{fig:psfsubtract1}.\label{fig:psfsubtract2}}
\end{figure*}

\begin{figure}[t]
\includegraphics[width=8.5cm]{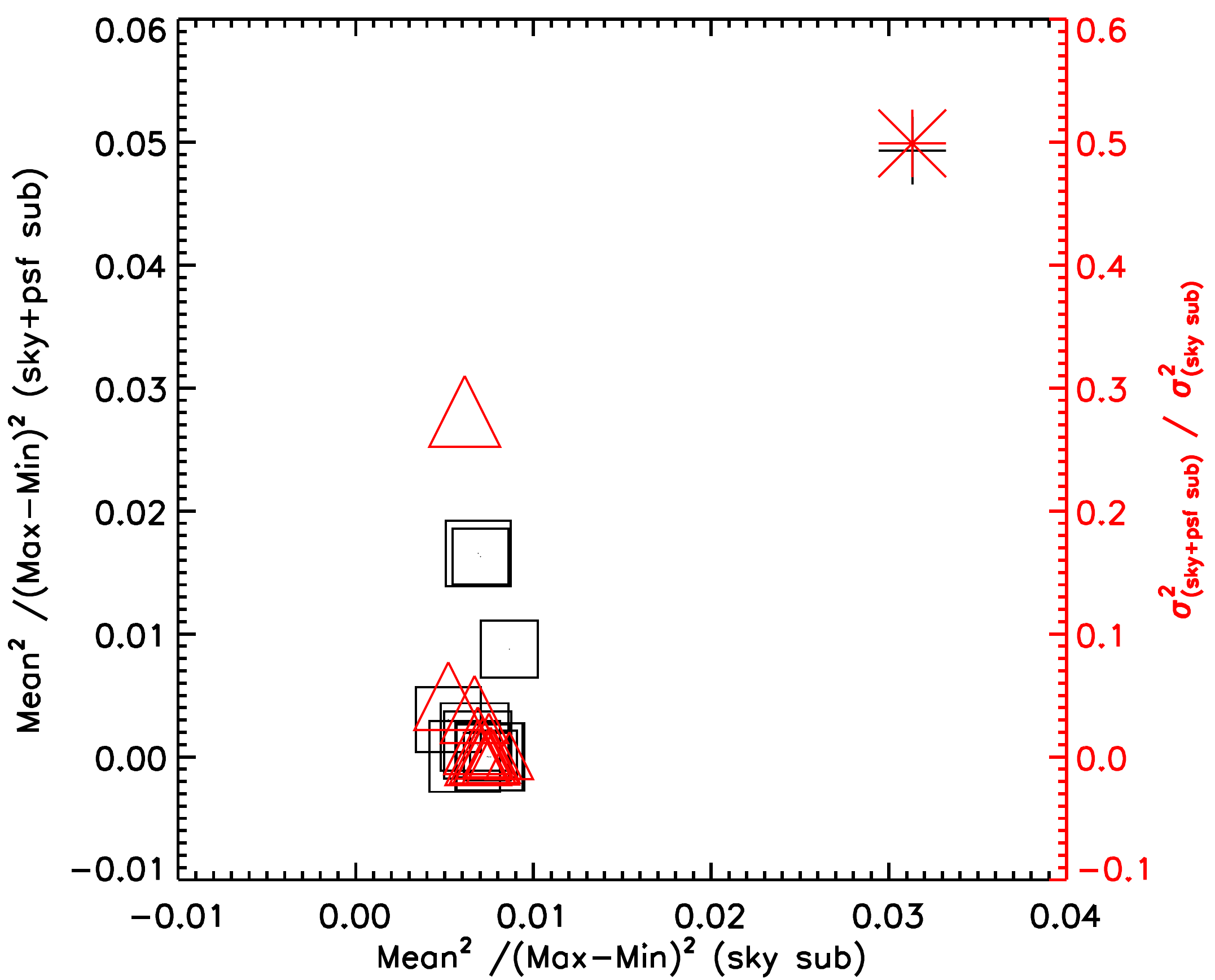}
\caption{Comparison of image statistics in the inner $0\farcs{7} \times 0\farcs{7}$ regions for the (bundle-stack) images with PSF subtraction (``psf+sky sub'') and without PSF subtraction (``sky sub'') for the reference sources (boxes and triangles) and the target (cross and asterisk symbols). The sizes of the symbols scale with the F160W magnitude. The brightest reference source (smallest symbol) has STmag=19.0, the faintest reference source has STmag=28.7. The latter is the object which is deviating from the other reference sources. Its PSF-subtracted image reveals a potential nearby faint neighbor.
 \label{fig:stat}}
\end{figure}

\end{document}